\documentclass[conference]{IEEEtran}

\usepackage{amssymb, amsthm}
\usepackage[cmex10]{amsmath}
\usepackage[tight,footnotesize]{subfigure}
\usepackage{url}
\usepackage[tight,footnotesize]{subfigure}
\usepackage{graphicx}
\usepackage{graphics}
\usepackage{amssymb}
\usepackage{amsmath}
\usepackage{amsthm}
\usepackage{color}
\usepackage{flushend}

\usepackage{multirow}
\usepackage{footnote}
\usepackage[bottom]{footmisc}

\theoremstyle{definition}
\newtheorem{thm}{Theorem}
\newtheorem{defn}{Definition}

\newtheorem{proposition}{Proposition}

\usepackage{tikz}

\usepackage[hang,small,bf]{caption}
\usepackage{tabulary}

\newcommand{\figref}[1]{\figurename~\ref{#1}}

\newcommand{\nop}[1]{}

\usepackage{lipsum}
\usepackage{xcolor}
\usepackage{soul}
\sethlcolor{black}
\makeatletter
\newif\if@blind
\@blindtrue %use \@blindfalse on final version

\if@blind \sethlcolor{black}\else

\fi
\usepackage{epstopdf}

\begin{document}

\title{ Service Modeling and Delay Analysis of Packet Delivery over a Wireless Link}	
%\title{ Modeling and delay performance analysis for Multi-layer Parameter Configuration of WSN Links}
%\title{Delay Performance for Multi-layer Parameter Configuration of WSN Links: Measurement and Analysis}

\author{\IEEEauthorblockN{ Yan Zhang$^{1}$,Yuming Jiang$^{1}$, Songwei Fu$^{2}$} %, Pedro Jos\'{e} Marr\'{o}n$^{2}$
	\IEEEauthorblockA{$^{1}$Department of Information Security and Communication Technology \\Norwegian University of Science and Technology (NTNU), Norway;\\
	$^{2}$Networked Embedded Systems, University of Duisburg-Essen, Germany	}}

\maketitle

\begin{abstract}
For delay analysis of packet delivery over a wireless link, several novel ideas are introduced. One is to construct an equivalent $G/G/1$ non-lossy queueing model to ease the analysis, enabled by exploiting empirical models of packet error rate, packet service time and packet loss rate obtained from measurement. The second is to exploit a classical queueing theory result to approximate the mean delay. For estimating the delay distribution, the newly developed stochastic network calculus (SNC) theory is made use of, forming the third idea. To enable this SNC based analysis, a stochastic service curve characterization of the link is introduced, relying on a packet service time model obtained from the empirical models. The focused link is a 802.15.4 wireless link. Extensive experimental investigation under a wide range of settings was conducted. The proposed ideas are validated with the experiment results. The validation confirms that the proposed approaches, integrating both empirical and analytical modes, are effective for service modeling and delay analysis. This suggests an integrated approach, now found previously, for quantitative understanding of the delay performance of packet delivery over a wireless link.

\end{abstract}

%\IEEEpeerreviewmaketitle

%\input{introduction-updated}
\section{Introduction}\label{sec:intro}
Many IoT (Internet of Things) applications are time-sensitive, requiring quality of service (QoS), particularly in terms of throughput, delay and loss guarantees, from the underlying wireless network. Due to the stochastic nature of wireless channel, it is challenging for a wireless link to reliably deliver data packets in real-time. %As an indication of the difficulty, URLLC (ultra-reliable low latency communications) has been identified as an important requirement and remains a big challenge for 5G and beyond. 
In the literature, a huge amount of research effort and results can be found for modeling and analyzing the performance of wireless links. The majority focuses on throughput-related performance metrics such as Shannon capacity \cite{Tse-book} where delay is not taken into account. 
For delay and loss analysis, queueing models have been adopted where the service process may be characterized by the throughput process or from some Markov characterization of the channel process \cite{EC} \cite{TWC-2013}. In these investigations, delay is typically attributed to traffic concurrency and queuing and loss due to  to buffer overflow, and to simplify analysis, queues are often assumed to have infinite buffer space where buffer overflow is approximated by the exceeding probability of the backlog higher than a threshold \cite{EC}. 
In real networks, however, there are other factors that can contribute to the delay and/or loss of a packet. They include finite block length coding-caused inherent error, other reasons-caused decoding errors, collision with other packets on the air, exceeding the allowed re-transmission times, etc. Taking all these factors into account can easily make the analysis formidably difficult or even intractable (e.g., see \cite{TOIT-2015} for an example analysis and \cite{Jiang-2015} for more discussion).  
%have been made to model and analyze the QoS performance of wireless links. 

Specifically, there are many existing theoretical analysis and modeling works that focus on delay guarantee analysis for packet delivery over wireless links, such as probabilistic delay bound analysis by applying the stochastic network calculus theory \cite{Jiang06, Jiang2008book, Jiang2010}, and combining models from Markov processes and queuing theory \cite{QS76}. Those delay analysis models require a number of simplifications and assumptions, e.g., knowing the distribution of the service time, which, however, may not always be satisfied or known in real-world wireless transmission. Hence, only relying on such analytical results to estimate the link performance is not enough.

There are also empirical measurement-based studies for packet delay and loss \cite{Ferrari:2007:WSN:1283620.1283661}, \cite{6424813}, \cite{5594493} and \cite{6662477}. However, most of them only consider one traffic pattern (e.g. periodic or Poisson) and some fixed stack parameter settings on the performance. Due to this, they usually lack deep analysis for the joint effect of multiple parameters on the delay or loss performance. As one step forward, in our previous works \cite{ICDCS2015, WCNC2018, PIMRC2018}, we made a set of extensive experimental studies for performance investigation of wireless links.  
%where, for instance, close to {\em 50 thousand} parameters configurations were considered and the meta-data of more than {\em 200 million} packets were collected in the work and the data set is publicly available \cite{due-packet-delivery-2015-03-30}. 
Based on the obtained experimental datasets, we have comprehensively investigated the effects of {\em typical parameters} from PHY, MAC and Application Layer on the link performance and presented several empirical models to quantify the joint effects of the stack parameters \cite{ICDCS2015, WCNC2018, PIMRC2018}. However, these models focus on throughput \cite{ICDCS2015} and packet loss rate \cite{WCNC2018, PIMRC2018}, leaving delay largely untouched. % \cite{experimentdataNES}

%Despite its importance, measuring, estimating or predicting the delay performance of a real-world wireless link is highly challenging. Firstly, due to the random effects of a wireless link, potential collisions, the limited number of maximal retransmission and the limited memory in the user nodes, a packet might be lost during the transmission or dropped during queueing. Meanwhile, varying packet transmission delay and queuing delay also lead to a rather complex stochastic characterization of the delay. Secondly, real world IoT network application conditions are highly diverse, often time-varying and traffic-varying, which additionally impacts the packet delay and loss performance. 

%It is impossible to come up with a set of experiment scenarios that are representative for all traffics. Therefore only relying on the empirical studies on off-line experimental testbeds to instruct the parameter configuration to meet the QoS performance guarantee in practical application, it is unreliable and inapplicable in real IoT applications. 
%To address those challenges, building upon the previous experimental studies \cite{ICDCS2015, WCNC2018, PIMRC2018}, 
To bridge the gap, we introduce in this paper a novel, integrated approach for service modeling and delay analysis of a wireless link, which combines empirical models obtained from the measurement study with and analytical models from the classical queueing theory \cite{QS76} and the stochastic network calculus (SNC) theory \cite{Jiang06, Jiang2008book, Jiang2010}. Specifically, for delay analysis, we propose a $G/G/1$ non-loss equivalent queuing model integrating measurement-based empirical models. Built upon the equivalent $G/G/1$ model, we explore an existing queueing analysis result for mean delay estimation. In addition, for delay distribution performance, we propose a SNC-based analytical approach. Central to this approach is the novel idea of finding the stochastic service curve characterization of the wireless link based on empirical models obtained from measurements. Both average delay and delay distribution models are validated with measurement data. The delay performance of the wireless link is further examined and discussed under different input traffic patterns. 

\nop{
The major contributions of this paper are summarized as follows:

\begin{itemize}
	\item 
	Based on the collected experimental data, We first establish some fundamental statistical empirical models, including Packet Error Rate (PER), packet loss rate mean and variance and packet deliver time, to quantify the partial performances which can be used for our following queuing analysis.
	
	\item For given traffic pattern in experiments, by applying queuing theory we propose a G/G/1 non-loss equivalent queuing model and validate it with the collected data to predict the link average delay performance. 
	
	\item 
	For general traffic arrival patterns, based on stochastic network calculus theory and empirical models we develop a practical delay prediction approach to characterize the packet delay distribution in wireless link. In this approach, We are first to introduce the empirical moment generation function (MGF) approximation model of packet service time to model the packet service process. Furthermore, by applying stochastic network calculus theory the delay performance analysis is conducted, where the delay bounds are shown. Lastly, the performance prediction method has been validated by our test-bed experiments in different traffic arrival. 
	
\end{itemize}

The rest of this paper is structured as follows. Section \ref{sec: experiment-setup} present our measurement
methodology and parameters configurations. Section \ref{sec:system-mod} presents the system overview and packet delivery scheme for WSN link. Section \ref{sec:emprical-mod} shows the measurement results on the basic statistics behaviors of packet link delivery. Section \ref{sec:analy-mod} introduces the equivalent G/G/1 non-loss queuing model and validates it. 
Section \ref{sec:analy-mod2} presents the stochastic performance bounds and validations for general traffic arrival. Finally, we conclude the paper with summary and future work in Section \ref{sec:conclusion}.
}

The rest of this paper is structured as follows. Section \ref{sec: experiment-setup} introduces our measurement
setup, parameter configurations, and packet delivery scheme. Section \ref{sec:emprical-mod} reports empirical models based on statistical analysis of measurement results. Section \ref{sec:analy-mod} introduces the non-loss $G/G/1$ queueing model and the classical queueing theory result for mean delay, where comparisons of analytical and measurement results are conducted to validate both the queueing model and the mean delay. Section \ref{sec:analy-mod2} introduces a SNC based approach to estimate the delay distribution, where validation under different traffic arrival patterns is also provided. Finally, we conclude the paper in Section \ref{sec:conclusion}.

\section{The Experiment Setup} \label{sec: experiment-setup}
\subsection{The wireless link and stack parameter configuration} 
\begin{figure}[th!]
	\vspace{-0.2cm}%
	\centering
	\includegraphics[width=1.0\columnwidth]{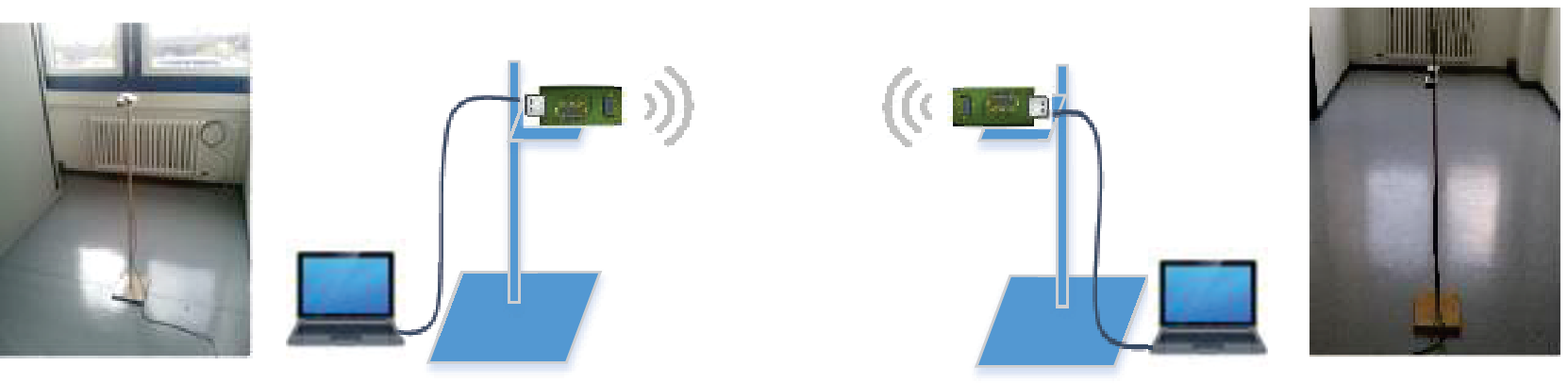}   \\
	\includegraphics[width=1.0\columnwidth]{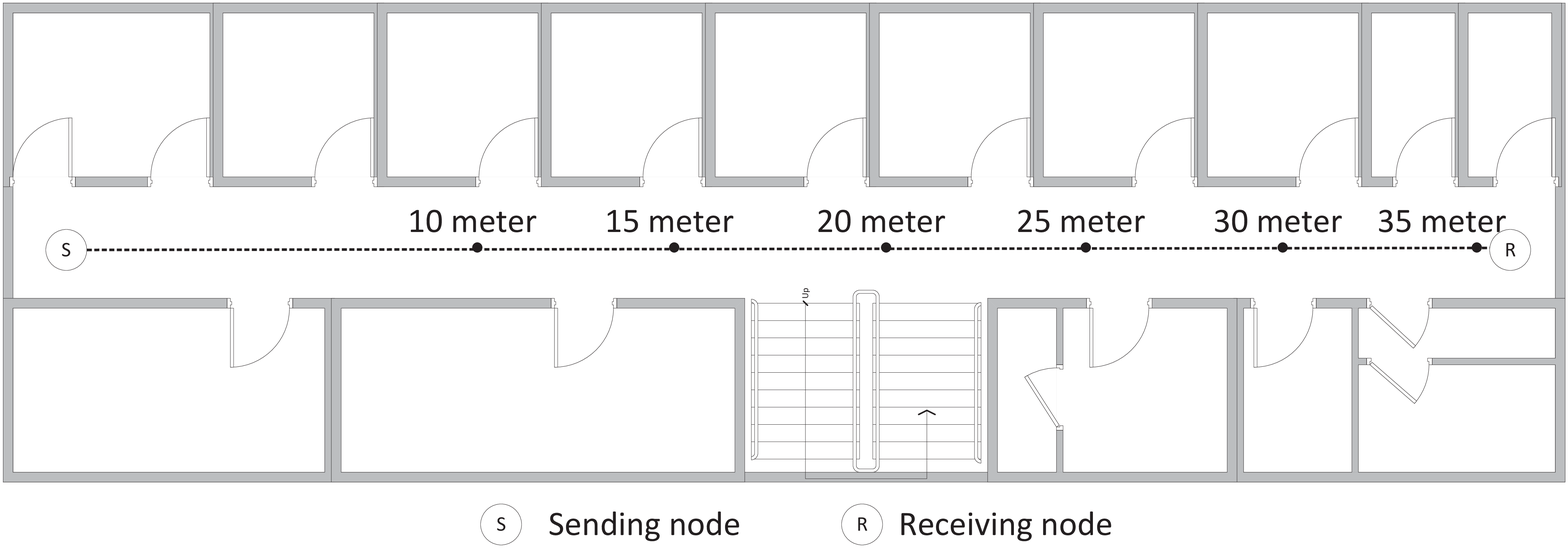} 
	\caption{Experiment setup}
	\vspace{-0.3cm}%
	\label{expFigure1}
\end{figure}

%In this section, in order to set the context, we briefly introduce our previous experimental studies for multi-layer parameter configuration of wireless links\cite{ICDCS2015}. % for service modeling and delay analysis
We consider a 802.15.4 link. To obtain an in-depth understanding of its service and delay performance, we conducted an extensive set of experiments in an indoor office building environment. We employed a sender-receiver pair of TelosB motes, each equipped with a TI CC2420 radio using the IEEE 802.15.4 stack implementation in TinyOS. As shown in ~\figref{expFigure1}, the experiments were conducted in a long hallway. Each mote was fixed on a wooden stand of 0.7-meter high and the positions of nodes. For each experiment, we maintained line-of-sight (LoS) between the two motes. During the experiments, university students and employees may walk in the hallway. 

In each experiment, the sender sends packets to the receiver under a particular stack parameter configuration. For each stack parameters configuration, 7 key parameters residing at different layers are considered. %Table~\ref{tab:ExpParameter} gives a summary of these parameters and their value ranges as well as the rationales behind the considered values. 
Specifically, at the PHY layer are the {\em distance} ($d$) between nodes and the {\em transmission power} level ($P_{\text{TX}}$). At the MAC layer are the \textit{maximum number of transmissions} ($N_{\text{maxTries}}$), the \textit{retry delay time} for a new retransmission ($D_{\text{retry}}$), and the \textit{maximum queue size} ($Q_{\text{max}}$) of the queue on top of the MAC layer used to buffer packets when they are waiting for (re-)transmission. At the Application layer are the \textit{packet inter-arrival time} ($T_{\text{pit}}$) and the \textit{packet payload size} ($l_D$). This setup is the same as used in our previous investigation \cite{ICDCS2015}, where more detailed information about these parameters' settings can be found. Different from \cite{ICDCS2015},  in the present paper, we are interested in service and delay characteristics of the link.   

%Different from \cite{ICDCS2015},  in the present paper, we are interested in system queuing behavior characteristics.  In total, close to {\em 50 thousand parameter configurations} were experimented and detailed transmission information of more than {\em 250 million packets} was collected, which provides statistical information for modeling the data delivery performance. 
 
%Based on some basic empirical models for measured periodic packet service process and our queuing analytical approach in following sections, we can obtain the delay performance for general traffic arrival. In order to evaluate our analysis results, we conduct additional experiments by giving two different packet arrival patterns which include Poisson and Markov On/Off traffic arrival. In total experiments, close to 50 thousand parameter configurations were experimented and detailed transmission information of more than 250 million packets was collected, which provides statistical information for modeling the data delivery performance. 

\nop{
\setcounter{footnote}{-1}

\begin{savenotes}
	\vspace{-0.3cm}%
	\begin{table*}[htbp]
		\centering
		\caption{Stack parameter configurations\cite{ICDCS2015}}
		\begin{tabulary}{0.95\textwidth}{|c|l|l|l|}
			\hline
			\textbf{Layer} & \textbf{Parameters} & \textbf{Parameter values}& \textbf{Rationale} \\ 
			\hline
			\textbf{Appl.} & $T_{pit}$: packet inter-arrival time (ms) & 10, 15, 20, 25, 30, 35, 40, 50 & Small $T_{pit}$ is selected to create high traffic arrival rate.\\
			%\hline			
			& $l_D$: packet payload size (bytes) & 20, 35, 50, 65, 80, 95, 110 & The maximum payload size for the used radio stack is 114 bytes.\\
			\hline
			\textbf{MAC} & $Q_{max}$: maximum queue size (packets) & 1, 30, 60 & The values represent no queue, medium and large queue.\\
			%\hline
			& $N_{maxTries}$: maximum \# of transmissions & 1, 3, 5 & 5 attempts are enough in our worst channel condition.\\
			%\hline
			& $D_{retry}$: retry delay (ms) & 30, 60 & Two different values are chosen to test the effect of retry delay.\\
			\hline
			\textbf{PHY}	& $P_{tx}$: transmission power level & 3,7,11,15,19,23,27,31 & TinyOS provides 8 output power levels from -25 dBm to 0 dBm. \\
			%		\textbf{PHY}	& $P_{TX}$: transmission power level & 1, 2, 3, 4, 5, 6, 7, 8 & TinyOS provides 8 output power levels from -25 dB to 0 dB. \\		%\hline
			& d: distance between nodes (meter) & 10, 15, 20, 25, 30, 35 & The maximum distance in the experiment scenario is 35 m.\\
			\hline
		\end{tabulary}		
		\label{tab:ExpParameter}
		\vspace{-0.5cm}%
	\end{table*}
\end{savenotes} 

\vspace{0.2cm}
To obtain an in-depth understanding of wireless link performance under different stack parameters configuration in WSNs, we conducted an extensive set of experiments in an indoor office building environment. We employed a sender-receiver pair of TelosB motes, each equipped with a TI CC2420 radio using the IEEE 802.15.4 stack implementation in TinyOS. As shown in ~\figref{expFigure1}, the experiments were conducted in a long hallway. Each mote was fixed on a wooden stand of 0.7-meter high and the positions of nodes. For each experiment, we maintained line-of-sight(LoS) between the two motes. During the experiments, university students and employees may walk in the hallway. In each experiment, the sender sends packets to the receiver under a particular stack parameter configuration. For each stack parameters configuration, 7 key parameters residing at different layers are considered. Table~\ref{tab:ExpParameter} gives a summary of these parameters and their value ranges as well as the rationales behind the considered values. 
%At the PHY layer are the {\em distance} ($d$) between nodes and the {\em transmission power} level ($P_{\text{TX}}$). At the MAC layer are the \textit{maximum number of transmissions} ($N_{\text{maxTries}}$), the \textit{retry delay time} for a new retransmission ($D_{\text{retry}}$), and the \textit{maximum queue size} ($Q_{\text{max}}$) of the queue on top of the MAC layer used to buffer packets when they are waiting for (re-)transmission. At the Application layer are the \textit{packet inter-arrival time} ($T_{\text{pit}}$) and the \textit{packet payload size} ($l_D$). 

Unlike previous works, in this paper we are interested in system queuing behavior characteristics.  Based on some basic empirical models for measured periodic packet service process and our queuing analytical approach in following sections, we can obtain the delay performance for general traffic arrival. In order to evaluate our analysis results, we conduct additional experiments by giving two different packet arrival patterns which include Poisson and Markov On/Off traffic arrival. In total experiments, close to 50 thousand parameter configurations were experimented and detailed transmission information of more than 250 million packets was collected, which provides statistical information for modeling the data delivery performance. 
}
%\input{systemmodel}

%\section{IEEE 802.15.4 overview and data delivery scheme in wireless link}\label{sec:system-mod}
\subsection{The data delivery scheme}\label{sec:system-mod}

%In this section, we will briefly introduce the IEEE 802.15.4 standard and its implementation in TinyOS in order to understand how QoS performances are affected by the interaction of the different features in the system. 

The IEEE 802.15.4 standard \cite{4040993} enables wireless connectivity between ultra-low power devices in wireless personal area networks (WPAN). The IEEE 802.15.4-standard both defines the physical layer and the medium access layer. Using the industrial scientific medical (ISM) band of 2.4 GHz, direct-sequence spread spectrum (DSSS) and phase shift keying (O-QPSK) modulation scheme at the PHY layer achieves a data rate of 250 kbps.The MAC layer defines two different channel access methods: beacon enabled and non-beacon enabled modes. In this paper, a non-beacon-enabled unslotted CSMA/CA mechanism is considered. The unslotted CSMA/CA procedure including backoff procedure and packet retransmission proceeds based on acknowledgments. Each time a generated data packet will wait for a random backoff time to check whether the channel is busy or not before to be transmitted. If the channel is idle during the backoff period, the device transmits its data packet. When the channel is busy, the random backoff procedure is repeated. When retransmissions are enabled, the destination node must send an acknowledgement (ACK) just after receiving a data frame, otherwise data frame will be retransmitted up to MACMAXFRAMERETRIES times, and then dropped. 
 
 \begin{figure}[htbp]
 	\centering
 	\includegraphics[width=1.0\columnwidth]{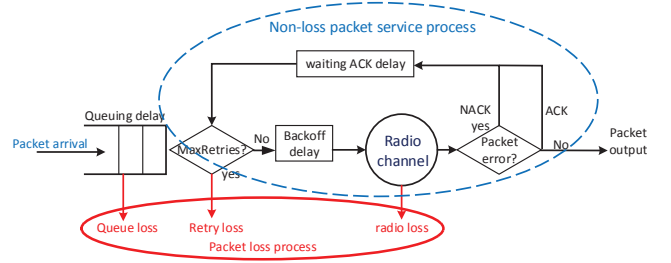}
 	\caption{Packet transmission system}% change the name
 	\label{sysmodel}
 \end{figure}

TinyOS \cite{tinyos}, an operating system for embedded wireless sensor networks, used to implement the actual data delivery is slightly different from the parameters specified in IEEE802.15.4. In our experimental study, the packet delivery scheme is implemented in TinyOS 2.1, as illustrated in \figref{sysmodel}. At the PHY layer, TinyOS allows the user to configure the radio transmit power to eight different levels from 3 to 31 (transmit power ranges from -25dB to 0dB) at compile time and during run-time. On top of the MAC layer, a buffer with maximum length is implemented to queue the application packets. Once a queue is full, newly arriving packets are dropped. The packets in the queue are served based on the FIFO (First In First Out) policy. Each time the sender node wishes to transmit data frames, it shall wait for a random backoff period (initialBackoff). If the channel is found to be idle, following the random backoff period, the device shall transmit its data. After the frame is sent, a copy of each transmitted packet is temporarily kept in a ``waiting buffer'' until the ACK of that packet is received. After a maxAckWaitDuration time, if NACK received, it triggers a retransmission until that the maximum number of retransmissions is reached and the packet is discarded. Between each retry the amount of delay time can be specified by TinyOS as retry delay. Once an ACK is received, the packet is removed from the waiting buffer and a new packet is transmitted. As \figref{sysmodel} shows, a packet loss may be due to queueing, retry and radio loss. The packet delay mainly includes transport and retry delays. The transport delay is subdivided into queueing, backoff, ACK duration and transmission delays. The transmission delay is defined as the time from a packet’s first generated until its successful arrival at the receiver (i.e., it includes all retransmission delays).

In total, close to {\em 50 thousand parameter configurations} were experimented and detailed transmission information of more than {\em 250 million packets} was collected, which provides statistical information for modeling the data delivery performance. A related dataset has been made publicly available \cite{due-packet-delivery-2015-03-30}. 

\section{ Measurement-Based Empirical Models }\label{sec:emprical-mod} %: First step  

%After a long term measurement experiment, we have collected large dataset for periodic packet arrival traffic pattern. In this section, we firstly quantitatively present several statistical empirical models for packet error rate, service delay time and  loss rate. Then later, those models will be applied in our analytical approach in other general packet arrival traffic pattern. All these statistical empirical models are reported with 95\% confidence level.

%
%There are several challenges to identify the impacting parameter configuration to the delay performance. The packet delivery perforamnce is highly influenced by variations of the randomness of channel environment.
%errors makes it even difficult to directly obtain packet delay performance results if we only depend on the experiment data analysis. To address those issues, in this section, we firstly quantitatively report several basic  statistical empirical models  and then we investigate those models later.

In this section, based on measurement results, statistical empirical models for packet error rate, service time and loss rate are presented. The measurement dataset for periodic packet arrival traffic pattern is used as the basis. %, where inter-packet time is so large that no queueing is found: This is to avoid queueing delay caused by traffic burstiness such that the measured per-packet delay can be considered as its service time by the link when adopting the queueing model as to be discussed in the next section. 
The obtained empirical models will be applied in later delay analysis. All these statistical results are reported with 95\% confidence level. %See \cite{ICDCS2015, WCNC2018, PIMRC2018} for more information about how to extract empirical models from similar experiment datasets. 

 \subsection{Packet error rate (PER)}
 \label{sec:per}
 
\begin{figure}[htbp]
   	\centering
   	\includegraphics[width=1.0\columnwidth]{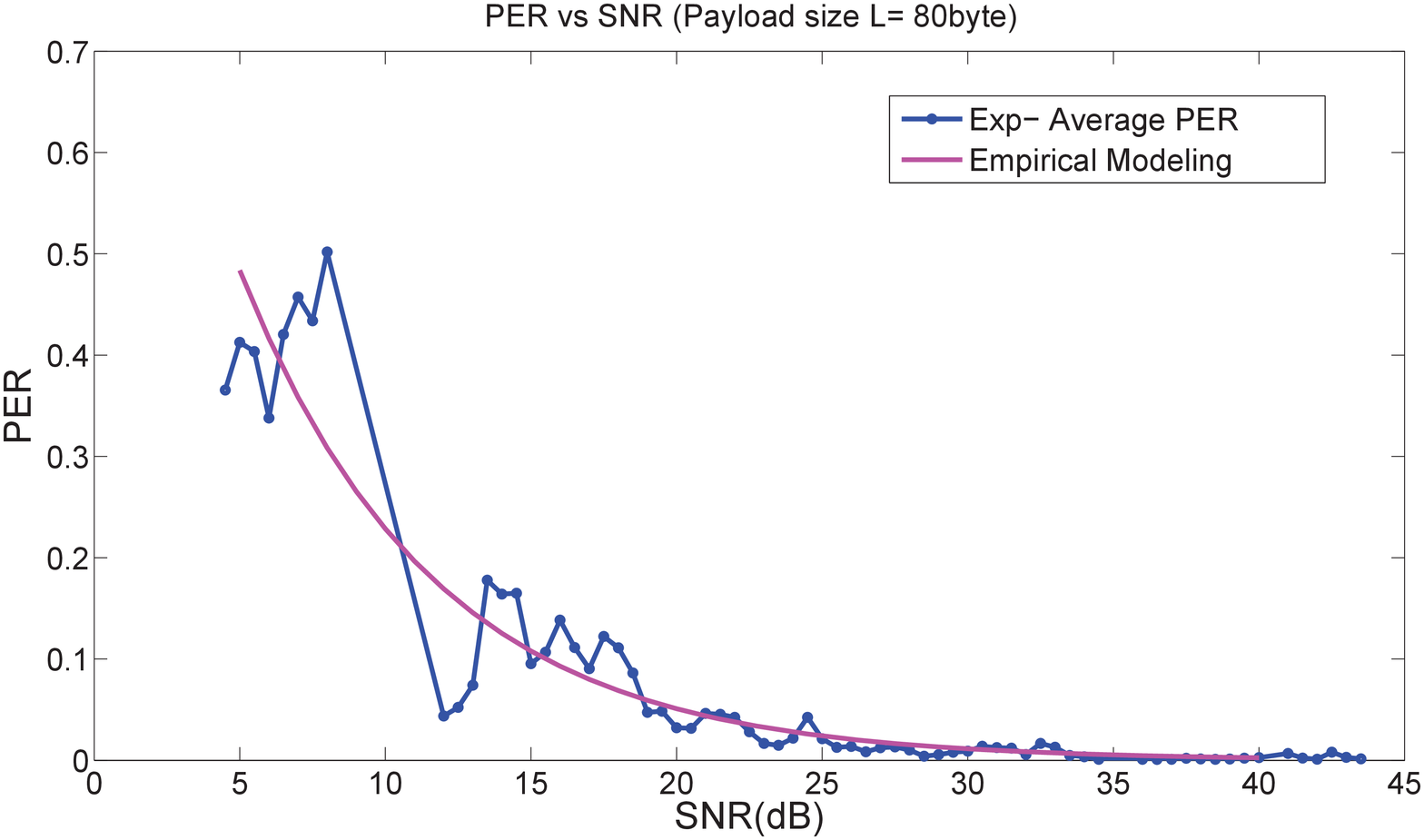}
   	\caption{Packet error rate}%{Modeling packet error rate}% change the name
   	\label{PERmodel}
   \end{figure}

PER is the ratio of the number of \nop{incorrectly received} unacknowledged data packets to the total number of transferred packets. By curve fitting over the measured average PER data as shown in \figref{PERmodel},  PER can be modeled as an exponential function of Single-to-Noise Ratio (SNR) and packet payload as: 
 \begin{equation} \label{equ:PER}
 P_e=\alpha\cdot l_D\cdot \exp(\beta\cdot \textrm{SNR}),
 \end{equation}
 with $\alpha=0.0128$ and $\beta=-0.15$ for our tested environment. %The fitting results are depicted in \figref{PERmodel}.

 \subsection{Packet service time} 
 \label{sec:Tservice}
 The packet service time, denoted as $T$, is defined as the time interval between when a packet is sent and when it is received by the receiver node. Specifically, the packet service time depends on (1) $\mathit{t_{SPI}}$ -- the one-time hardware SPI bus loading time of a data frame; (2) $\mathit{t_{frame}}$ -- the time to transmit a frame consisting of packet payload and overhead; (3)  $\mathit{t_{MAC}}$ -- MAC layer delay consisting of two parts: $t_{TR}$ and $t_{BO}$, where $t_{TR}$ is the turn around time (which is set to be $0.224ms$ in our experiments) and $t_{BO}$ is the average value of initial backoff period (which is set to be $5.28ms$); (4) $\mathit{t_{ACK}}$ -- the ACK frame transmission time if ACK frame is received, and based on prior tests $t_{ACK}\approx 1.96 ms$; (5) $\mathit{t_{waitACK}}$ -- the maximum software ACK waiting period (set to be $8.192ms$); (6) $\mathit{n_{tries}}$ -- the number of transmissions to deliver each packet; (7) $\mathit{D_{retry}}$ -- the delay between two consecutive retransmissions. 

There are two cases in per-packet service time, denoted as $T_{ACK}$ and $T_{nonACK}$ respectively, depending on whether a packet is successfully transmitted (i.e., ACK received): 
  \begin{itemize}
  	\item If $1\leq n_{tries}\leq N_{maxTries}$ (with ACK),  
  	\begin{equation}\label{equ:servicetimeACK}
  	T_{ACK} = t_{SPI}+t_{succ}+(n_{tries}-1)\cdot t_{retry}
  	\end{equation}
  	\item If $n_{tries}= N_{maxTries}$ (with non-ACK), 
  	\begin{equation}\label{equ:servicetimenonACK}
  	T_{nonACK}= t_{SPI}+t_{fail}+(N_{maxTries}-1)\cdot t_{retry}
  	\end{equation}
  \end{itemize}
  where
  \begin{eqnarray}
  	t_{succ}&=&t_{MAC}+t_{frame}+t_{ACK}\nonumber\\%\label{equ:Tsucc} \\
  	t_{fail}&=&t_{MAC}+t_{frame}+t_{waitACK}\nonumber\\%\label{equ:Tfail}\\
  	t_{retry}&=&D_{retry}+T_{fail}\nonumber%\label{equ:Tretry}
  \end{eqnarray}
 
   \begin{figure}[th]
  	\centering
  	\includegraphics[width=1.0\columnwidth]{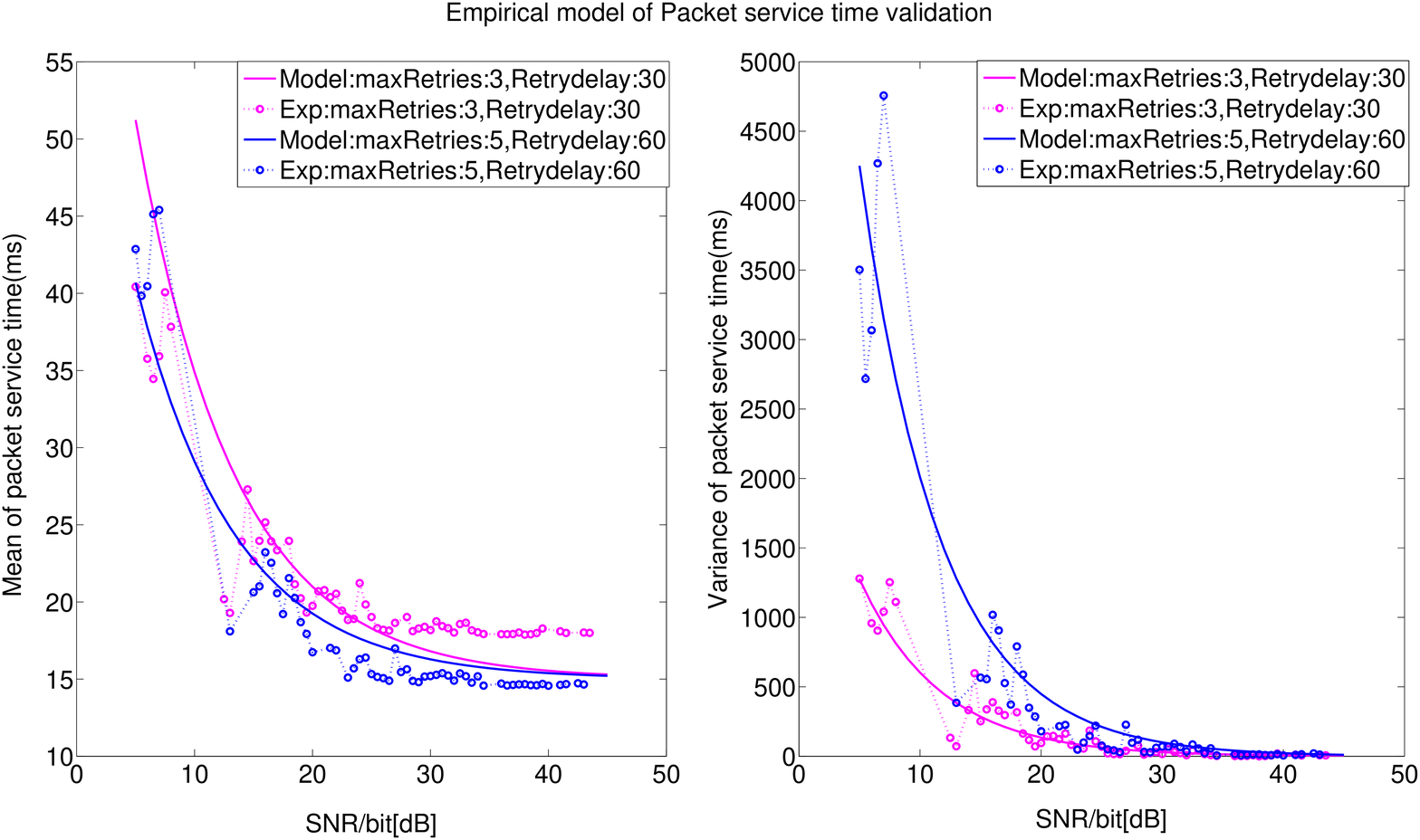}
  	\caption{Packet service time: mean and variance}% change the name
  	\label{Tservicemodel}
  \end{figure}
  
Due to the random nature of wireless transmission, the actual numbers of retransmission of a packet $n_{tries}$ is a random value. Hence the packet service time $T$ is also random. From the measurement results, we found that the mean and variance of packet service time, denoted as $E(T)$ and $Var(T)$ respectively, are sensitive to the maximum number of transmissions $N_{maxTries}$, payload size $l_D$, the delay between two consecutive retransmissions $D_{\text{retry}}$ and $\text{SNR}$. Through curve fitting, empirical models for them are
 \begin{eqnarray} 
	E(T) & = & \frac{0.06}{n_{\text{maxTries}}} D_{\text{retry}} l_D \exp(-0.12 \text{SNR})+15 \label{servicetime-mean} \\
        Var(T) & =& 30 N_{\text{maxTries}} D_{\text{retry}}\exp(-0.15 \text{SNR}) \label{servicetime-var} 
\end{eqnarray}
The fitting results are depicted in \figref{Tservicemodel}.

%If $N_{maxTries}=1$,
%\begin{align}\label{servicetime1}
%& & E(T_{service})= 0.1\cdot l_D\exp(-0.01 SNR)+9.9,\\
%& & Var(T_{service}) =17.5\exp(-0.11 SNR);
%\end{align}
\nop{	 	
%If $N_{maxTries}>1$, 
\begin{align}
	\label{servicetime2} &  & E(T) =\frac{0.06}{n_{\text{maxTries}}} D_{\text{retry}} l_D \exp(-0.12 \text{SNR})+15, \\
	                     &  & Var(T)=30 N_{\text{maxTries}} D_{\text{retry}}\exp(-0.15 \text{SNR});
\end{align}
where $E(T)$ and $Var(T)$ denotes the mean and the variance of packet service time. 
}

\subsection{Packet loss rate (PLR)}% for periodic packet arrival}

Packet loss over a wireless link is mainly caused by poor link quality and limited buffer size. The former may result in the drop of a packet after the maximum transmission attempts $N_{\text{maxTries}}$ have been tried. The latter will cause a packet being dropped when its arrival sees a full buffer. Note that PLR differs from PER. The difference is that while PER considers all transmissions including retransmissions, PLR only count it one time for each data packet no matter whether it may have been retransmitted multiple times.  

\begin{figure}[htbp]
  	\centering
  	\includegraphics[width=1.0\columnwidth]{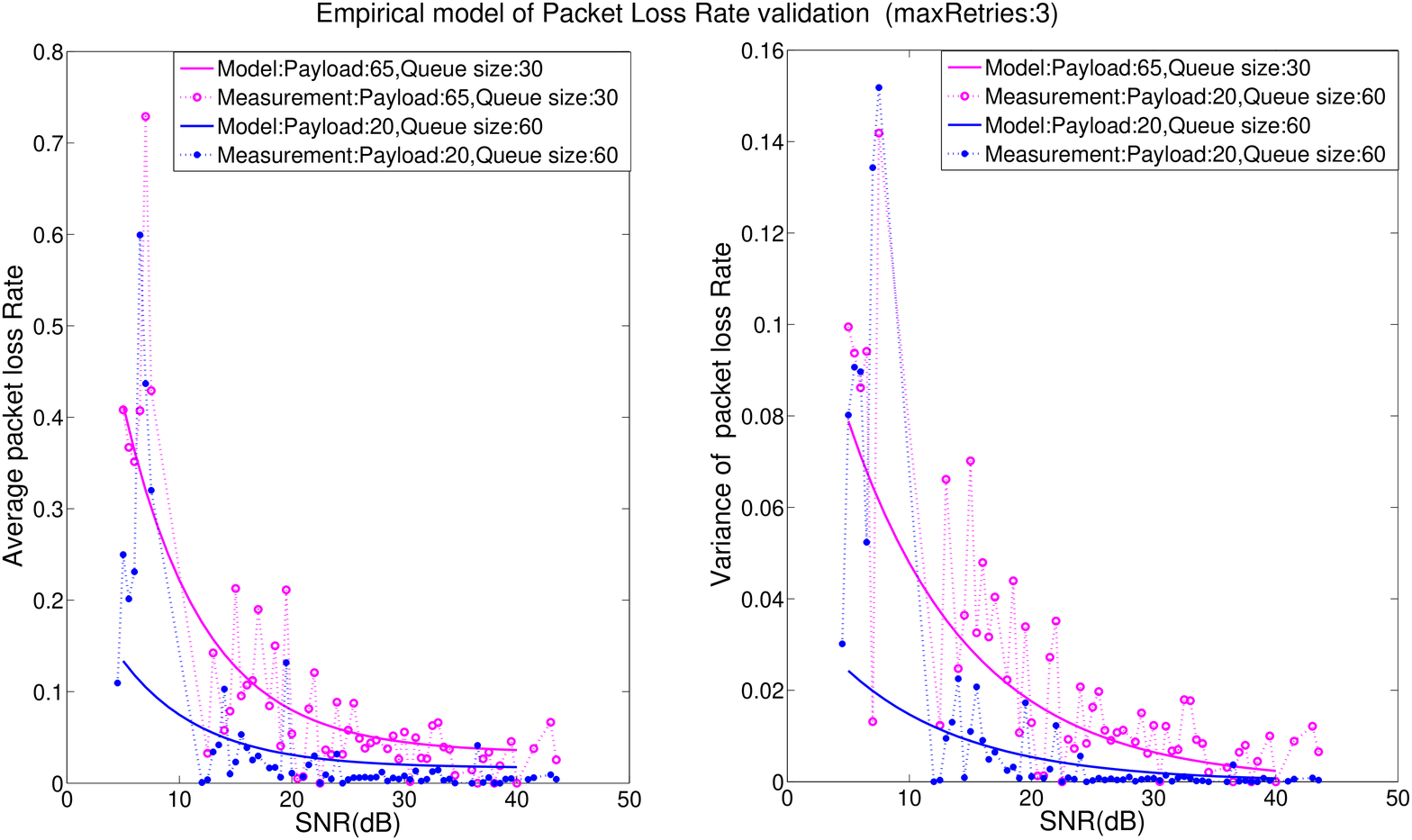}
  	\caption{Modeling packet loss rate }% change the name
  	\label{PLRmodel}
  \end{figure}

For PLR, we also consider its mean and variance, denoted as $E(\text{PLR}$ and $Var(\text{PLR})$ respectively. It is found that they are functions of payload size $l_D$, $\text{SNR}$, buffer length $L_q$, and maximum transmission attempts $N_{\text{maxTries}}$. By curve fitting as shown in \figref{PLRmodel}, the mean and variance of packet loss rate can be approximately modeled as
%Based on the analysis of our indoor experimental results, the Packet loss in wireless link can be mainly included by radio loss due to the poor link quality and queuing loss due to the imitation of queue length. To cope with more insights about the packets losses, based on our main experiment results study for periodic packet arrival pattern we model the mean and variance of packet loss rate ($E(\text{PLR})$ and $Var(\text{PLR})$) as a function of payload size$l_D$, $\text{SNR}$, queue length($L_q$) and maximum transmission attempts $N_{\text{maxTries}}$. The mean and variance of packet loss rate can be approximately modeled as the exponential functions as follows:
\begin{eqnarray}\label{equ:PLRmodels}
E(\text{PLR}) &=& \frac{l_D}{100}\exp(-0.14\text{SNR})+\frac{1}{L_q}\\
Var(\text{PLR}) &=&  \frac{l_D}{500}\exp(-0.1\text{SNR})
\end{eqnarray}

%The figure \figref{PLRmodel} shows that the modeling approximation of the mean and variance of packet loss rate are validated by our indoor experimental results.

%\begin{table*}[ht]
%	\centering
%	\caption{Models of Packe loss rate}\label{tab:PLRmodels}
%	\vspace{-0.2cm}%
%	\begin{tabular}{|c|c|}% total 3 colum
%		
%		\hline
%		$L_q=1$ & $L_q>1$ \\
%		\hline
%		\multicolumn{2}{|c|}{$N_{maxTries}=1$(no retransmission)}\\		
%		\hline		
%		$E(PLR)=\frac{l_D}{117}\exp(-0.06SNR)$\stepcounter{equation}\thetag{\theequation}&$E(PLR)= \frac{l_D}{95}\exp(-0.06SNR)$\stepcounter{equation}\thetag{\theequation}\\
%		
%		\hline
%		\multicolumn{2}{|c|}{$Var(PLR)= \frac{l_D}{1658}\exp(-0.04SNR)+0.02$\stepcounter{equation}\thetag{\theequation}}\\
%		\hline
%		\hline
%		\multicolumn{2}{|c|}{$N_{maxTries}>1$(with retransmission)}\\
%		\hline
%		$E(PLR)= \frac{l_D}{85}\exp(-0.14SNR)+\frac{1}{L_q}$\stepcounter{equation}\thetag{\theequation}
%		& $E(PLR)= \frac{l_D}{100}\exp(-0.14SNR)+\frac{1}{L_q}$\stepcounter{equation}\thetag{\theequation} \\
%		\hline
%		\multicolumn{2}{|c|}{$Var(PLR)= \frac{l_D}{500}\exp(-0.1SNR)$\stepcounter{equation}\thetag{\theequation}}\\
%		\hline		
%	\end{tabular}
%	\vspace{-0.5cm}%
%\end{table*}
%

%\input{analysis}
\section{Queueing Model for Delay Analysis}\label{sec:analy-mod}

In comparison to estimating packet loss rate, which can be easily done by counting the numbers of packets sent and received, estimating delay particularly its distribution tail or delay quantile, which is crucial for time-critical applications, is more challenging. This is not only because measuring delay requires time-synchronisation among nodes in the system and accurate time information from both the sender and the receiver, but also because estimating the delay distribution tail requires much higher number of packets to be involved and the tail can vary significantly with respect to different traffic patterns. 

To address this difficulty, our approach is to construct a queueing model on which delay analysis can be performed, including finding the delay tail distribution. In this section, we introduce the queueing model and validate the model using mean delay. In the next section, we further introduce how to estimate the delay tail and compare the analytical such results with measurement results under different traffic patterns. 

%In previous section, our measurement and empirical modeling study characterize the packet service delay(service time) in a periodic arrival traffic pattern. The packet delay in WSNs link consists of two components: queuing delay(waiting delay) and service delay(transmission delay). While queuing delay is highly variable and may be accompanied by variations of different parameters such as link quality, arrival traffic, queue size, numbers of retransmission etc. However, based on our measurement results it is difficult to address the queuing delay directly. we propose a equivalent G/G/1 non-loss queuing analysis model consisting empirical models for a wirelesss link to approximate the average delay performance.

%\subsection{Equivalent G/G/1 Queueing Model and delay approximation}
\subsection{Equivalent queueing model}

For a system as shown in \figref{sysmodel}, while looking simple, it is actually highly difficult or even intractable to conduct its delay analysis (e.g., see \cite{TOIT-2015} for an example analysis and \cite{Jiang-2015} for more discussion).  

\begin{figure}[thb]
	\centering
	\includegraphics[width=1.0\columnwidth]{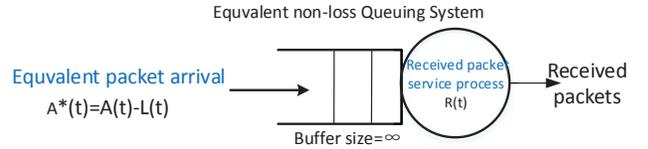}
	\caption{Equivalent non-loss queueing modeling  }% change the name
	\label{equmodel}
\end{figure}

In order to simplify the analysis we introduce an equivalent queueing model. The basic idea is to treat the system as a blackbox from the receiver view, where all received packets have been successfully delivered without the lost packets. As such, we model the transmission process of successfully received packets over the wireless link as a $G/G/1/\infty$ non-loss queuing system as shown in ~\figref{equmodel}.

Specifically, as illustrated in ~\figref{equmodel}, the (successfully received) packets arriving to the equivalent system are served in the FIFO manner with an infinite size buffer for possible queueing. Let $A(t)$ and $L(t)$ respectively represent the original packet arrival process and the loss process. Then, the packet arrival process after considering lost packets becomes $A(t)-L(t) \equiv A^{\ast}(t)$. In addition, we let $R(t)$ represent the service time process of successfully received packets.

%Directly using $G/G/1$ queuing model to obtain the queuing delay in our queuing system\ref{sysmodel} is difficult by the fact that the data loss or blocking is a frequent event when data deliver over an unreliable link as well as the queue length and the number of retries are limited. It is complicated to capture the packet service process in such loss retransmission queuing system. In order to simplify the system we offer an equivalent queuing model. The basic idea is to model a equivalent queuing system in the view of receiver part in which the packets have been successfully delivered without including the blocked and lose packets. It means the packet transmission process between two sensor nodes can be modeled as a G/G/1/$\infty$ non-loss equivalent queuing model shown in ~\figref{equmodel}.

%Specifically, as illustrated in ~\figref{equmodel}, the packet arrives into the equivalent system will be stored in the corresponding buffer(queue) in a First-In-First-Out (FIFO) policy, where the buffer size is infinite. Let $A(t)$,$R(t)$ and $L(t)$ respectively represent the packet arrival, the service time and loss process. We treat the whole transmission process including retransmission processes as a black box. The equivalent packet arrival process after considering lost packets is represented by $A^{\ast}(t)=A(t)-L(t)$. Due to the independence of packet arrival process and packet loss process, the mean and variance of the equivalent packet arrival rate($\lambda$ and $Var(A^{\ast})$) is obtained as:

%Due to the independence of packet arrival process and packet loss process, 
Given the mean and variance of packet loss rate (PLR), the mean and variance of the equivalent packet arrival rate, respectively denoted as $\lambda$ and $Var(A^{\ast})$, can be found as:
   \begin{eqnarray}
   \lambda &=& (\frac{1}{T_{int}})\cdot(1-E(PLR))   \label{equ:equarrivalmean} \\
     Var(A^{\ast}) &=& (\frac{1}{T_{int}})^2\cdot Var(PLR) \label{equ:equarrivalvar}
   \end{eqnarray}
  where $T_{int}$ denotes average packet inter-arrival time of the original packet arrival process. 
\nop{	   
   \begin{equation}\label{equ:equarrivalmean}
   \lambda=(\frac{1}{T_{int}})\cdot(1-E(PLR))   
   \end{equation}
   
  \begin{equation}\label{equ:equarrivalvar}
  Var(A^{\ast})=(\frac{1}{T_{int}})^2\cdot Var(PLR)   
  \end{equation}
  where $T_{int}$ is packet inter-arrival time.
  }
 
%\subsection{Delay approximation} 
\subsection{Model validation using mean delay}\label{sec-mvm} 

For a $G/G/1/\infty$ system that is not overloaded and hence has finite waiting time, let $\lambda$ denote the mean arrival rate and $\sigma^2_{A^{\ast}}$ its variance, and $T$ the mean service time and $\sigma^2_{T}$ its variance. The following approximation for waiting time in queue $ W_{q}$ is from the queueing theory, e.g. see \cite{QS76}:
 \begin{equation} \label{equ:GG1}
  W_{q}\approx \frac{\lambda(\sigma^2_{A^{\ast}}+\sigma^2_{T})}{2(1-\rho)}
 \end{equation}
where  $\rho=\lambda\cdot E(T)$ is the traffic intensity of the system. With (\ref{equ:GG1}), the mean system delay is simply:
\begin{equation}\label{equ:totaldelay}
      Delay = W_{q}+E(T)
\end{equation}   

For the equivalent $G/G/1/\infty$ system, $\lambda$ is the equivalent average packet arrival rate that can be found from (\ref{equ:equarrivalmean}) and $\sigma^2_{A^{\ast}}$ its variance from (\ref{equ:equarrivalvar}), and $E(T)$ and $\sigma^2_{T}$ can be found from  (\ref{servicetime-mean}) and  (\ref{servicetime-var}) respectively. With these, the mean delay can be analytically estimated from (\ref{equ:totaldelay}). 
 
\figref{gg1delay} compares mean delay results obtained through experiment and through the analytical model above under different parameter settings. It can be observed that a good agreement between the two results can be found in most cases. This confirms the effectiveness of the proposed equivalent $G/G/1/\infty$ and its validity for mean delay analysis. For the delay from measurement results has larger variation in small SNR region, we remark that this is due to lack of delay samples in this SNR region under our experiment environment. %However, due to the correlation between packet loss and inter-arrival traffic, our equivalent G/G/1 delay model can only good approximately predict the packet delay in periodic traffic inter-arrival pattern. Therefor, without the packet loss empirical model we can not obtain the delay prediction so far. To address this problem, we present the delay model of stochastic network calculus in the subsequent Sections.
  
 \begin{figure}[thb]
 	\centering
 	\includegraphics[width=1.\columnwidth]{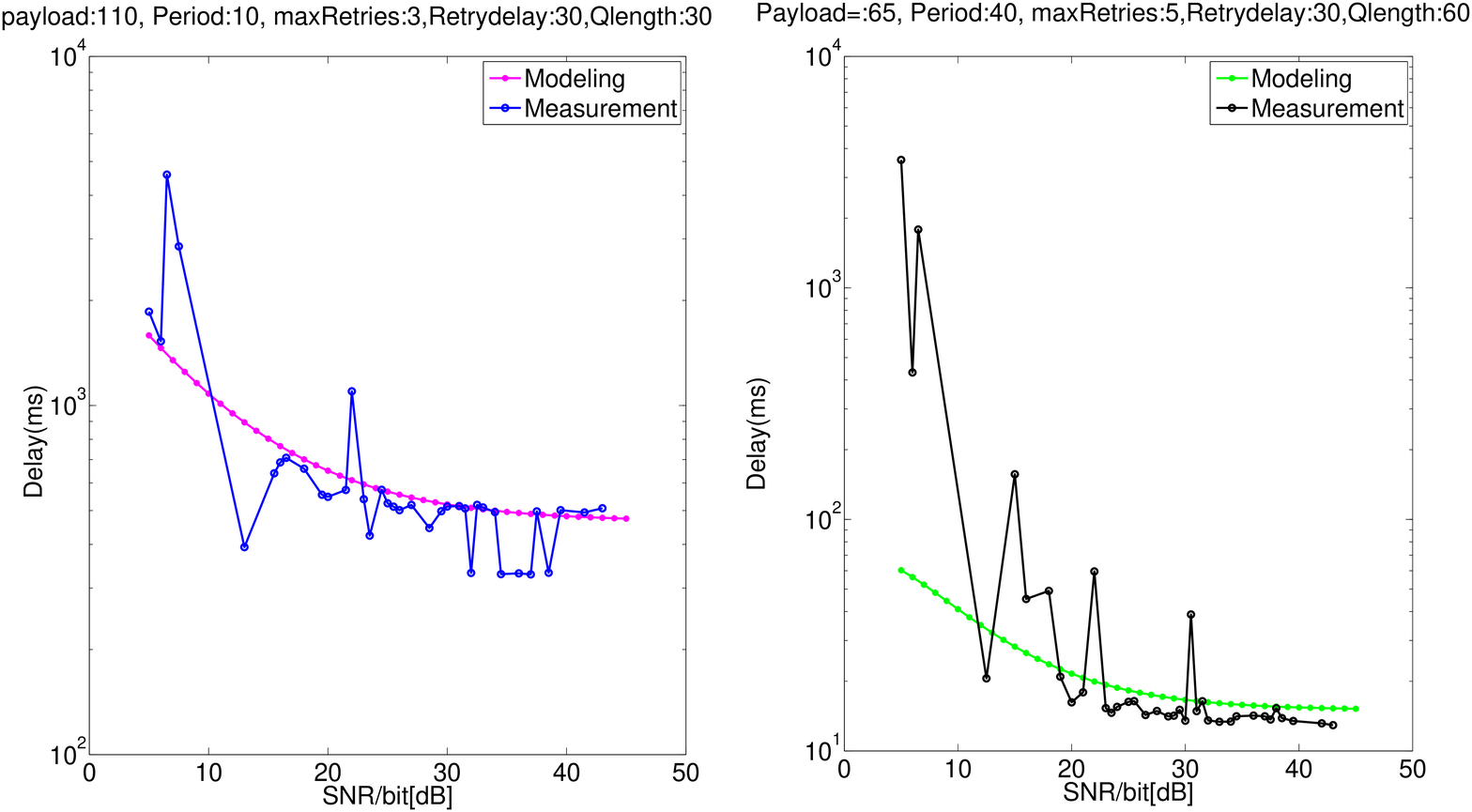}
 	\caption{Model validation: mean delay}% {Delay validation for periodic packet arrival based on G/G/1 queuing model. Model prediction and experiments results}change the name
 	\label{gg1delay}
 \end{figure}

\nop{
Based on the $G/G/1/\infty$ queue the approximation of queuing waiting time(queuing delay) $W_{q}$ in our equivalent system could be given as \cite{QS76}:
 \begin{equation} \label{equ:GG1}
  W_{q}\approx \frac{\lambda(\sigma^2_{A^{\ast}}+\sigma^2_{T_{service}})}{2(1-\rho)}
 \end{equation}
 where symbol $A^{\ast}$ and T denote the equivalent arrival interval and the service time, $\sigma^2_{A^{\ast}}$ and $\sigma^2_{T}$ denote the variances of them respectively\ref{equ:equarrivalvar}\ref{sec:Tservice}. $\lambda$ is the equivalent average packet arrival rate\ref{equ:equarrivalmean} and $\rho=\lambda\cdot E(T_{service})$ is the traffic intensity of the system\ref{equ:equarrivalmean}\ref{sec:Tservice}. The inequality \ref{equ:GG1} becomes equality when $\rho$ approaches 1. If $\rho\geq 1$, the system will be unstable and the queue will be full, the queuing waiting time can be approximately calculated as 
     \begin{equation}\label{equ:quedelayunstable}
     W_{q}\approx L_q\cdot E(T_{service})  
    \end{equation}
where $L_q$ is the queue length.

Considering the packet service delay$E(T_{service})$ the total packet delay of single wireless link can be approximately given as:

      \begin{equation}\label{equ:totaldelay}
      Delay\approx W_{q}+E(T_{service})
      \end{equation}   
          
 \subsection{Model Validation}
 The figure ~\figref{gg1delay} shows that the modeling results of the packet delay approximation is validated by our indoor experiments .

 \begin{figure}[thb]
 	\centering
 	\includegraphics[width=1.\columnwidth]{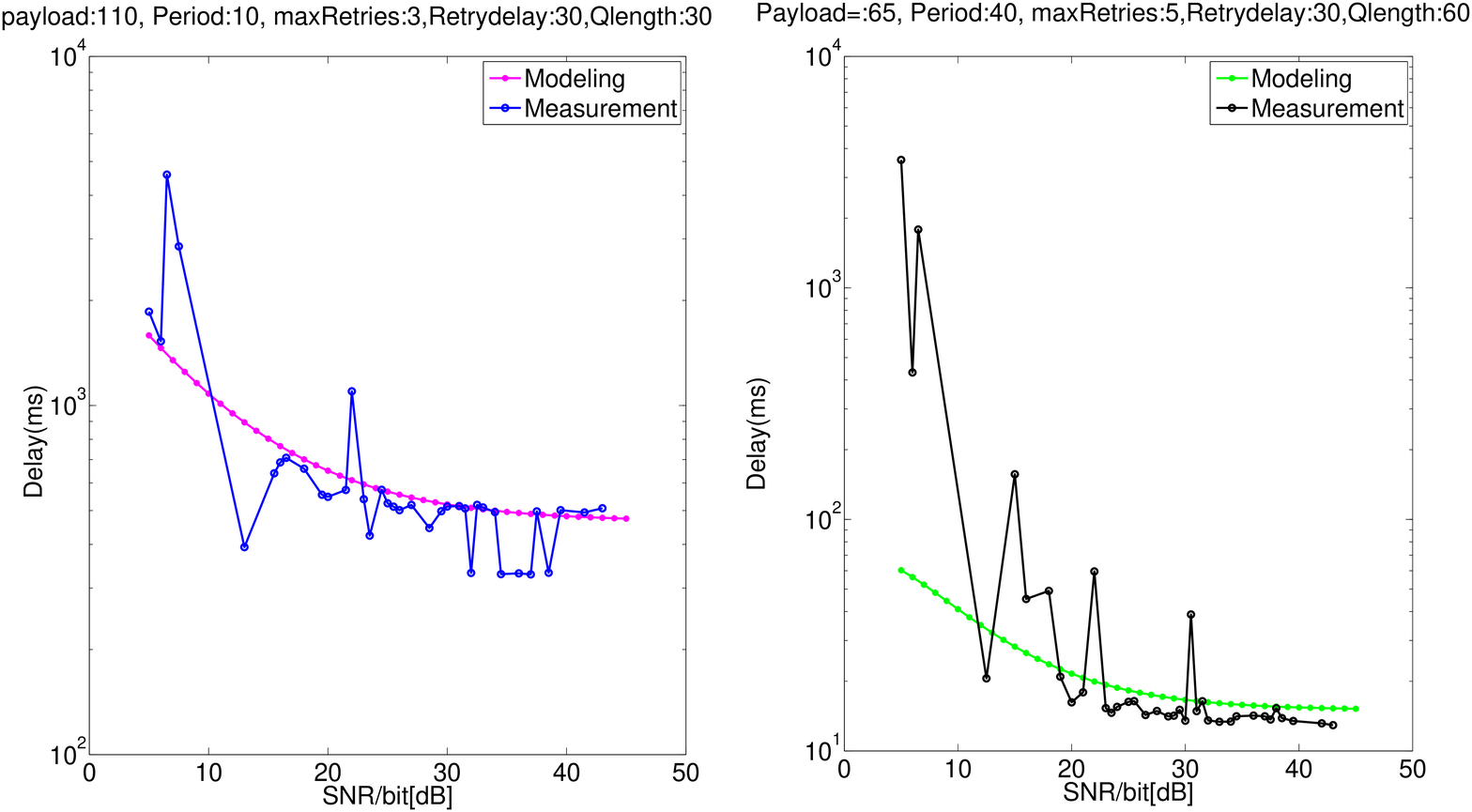}
 	\caption{Model validation: mean delay}% {Delay validation for periodic packet arrival based on G/G/1 queuing model. Model prediction and experiments results}change the name
 	\label{gg1delay}
 \end{figure}

 In \figref{gg1delay}, the approximation delay prediction for periodic traffic arrival is shown. The comparison between results obtained through experiment and through the model prediction under different parameters setting are reported. It can be observed that a good agreement between the two results can be found in most cases. We notice that the delay from measurement results is larger variation when SNR value is lower 10dB. This is due to that there are not enough large numbers of delay sample in SNR=5dB under our indoor experiment environment. It may lead to the average delay has large variation. However, due to the correlation between packet loss and inter-arrival traffic, our equivalent G/G/1 delay model can only good approximately predict the packet delay in periodic traffic inter-arrival pattern. Therefor, without the packet loss empirical model we can not obtain the delay prediction so far. To address this problem, we present the delay model of stochastic network calculus in the subsequent Sections.
} 

%\input{SNCanalysis}
%\section{ Performance Analysis in Stochastic Network Calculus}\label{sec:analy-mod2}
\section{Investigation on Delay Distribution}\label{sec:analy-mod2}

For applications requiring reliable real time data packet delivery, average delay is often not enough and the concern is more on the delay distribution, or more specifically the tail of the distribution, i.e., $P\{Delay > x\}$. However, for delay distribution of a $G/G/1/\infty$ system, there are few results from the classical queueing theory \cite{QS76}. In the following, a new queueing theory, known as stochastic network calculus \cite{Jiang06, Jiang2008book}, is resorted.

%Through our proposed equivalent G/G/1 non-loss analysis approach in Sec.\ref{sec:analy-mod}, we can predict the average delay under given traffic pattern in our experiments. To obtain delay prediction for other traffic pattern, in this section, different from previous work we propose a practical method for stochastic QoS analysis of the packet service process and find the probabilistic delay bound for general traffic arrival.

\subsection{Stochastic network calculus background}\label{sec:SNCthm}
%\subsection{Stochastic Network Calculus and performance analysis}\label{sec:SNCthm}
Stochastic network calculus is a queuing theory for stochastic performance guarantee analysis of communication networks \cite{Jiang06, Jiang2008book}. It is built upon two fundamental concepts, which are stochastic arrival curve (SAC) for traffic modeling and stochastic service curve (SSC) for server modeling \cite{Jiang2008book}. There are several definition variations of SAC and SSC. In this paper, the following are adopted \cite{Jiang2008book}. %In this paper, the virtual backlog property SAC is addressed, which explores the virtual backlog property of deterministic arrival curve $\alpha(t)$. It is defined as\cite{Jiang2008book}:

\begin{defn}(Stochastic Arrival Curve)\label{SSA}
A traffic flow is said to have a stochastic arrival curve $\alpha$ with bounding function $f$, if its arrival process $A(t)$ satisfies, for any $t\geq0$ and all $x\geq0$ there holds:
\begin{equation}\label{vbc SAC}
\Pr\{\sup_{0\leq s \leq t}[A(s,t)-\alpha(t-s)]> x\}\leq f(x).
\end{equation}
where $A(s,t)$ denotes the cumulative amount of traffic of the flow from time $s$ to time $t$, $A(t)= A(0,t)$.
\end{defn}

\begin{defn}(Stochastic Service Curve)\label{SSC}
A system is said to have a stochastic service curve $\beta$ with bounding function $g$, if for all $t\geq 0$ and all $x \geq 0$ there holds:
\begin{equation}\label{ssc}
\Pr\{A\otimes\beta(t)-O(t)> x\}\leq g(x),
\end{equation}
where $A \otimes \beta (t) \equiv in{f_{0 \le s \le t}}\left\{ {A(s) + \beta (t - s)} \right\}$, ${O}(t)$ denotes the cumulative output traffic amount up to time $t$. 
\end{defn}

\begin{thm}(Stochastic delay bound)\label{SNCbounds}
For a $G/G/1/\infty$ system, if the input has a stochastic arrival curve and the system provides a stochastic service curve to the input, as defined above respectively, and the arrival process is independent of the service process, then for the delay, the following bound holds \cite{Jiang2008book}:
\begin{equation}\label{delaybound}
P\left\{ {D(t) > h(\alpha  + x,\beta )} \right\} \le 1 - \overline f  *  \overline g (x)
\end{equation}
where$\overline f (x) = 1 - {\left[ {f(x)} \right]_1}$, $\overline g(x) = 1 - {\left[ {g(x)} \right]_1}$ with $[\cdot]_1$denotes $\max(\min(\cdot,1),0)$, $h(\alpha  + x,\beta)$ represents the maximum horizontal distance between curves $\alpha  + x$ and $\beta$, and $*$ is the Stieltjes convolution operation $\overline f  * \overline g (x) = \int\limits_0^x {\overline g } (x - y)d\overline f (y)$.
\end{thm}

In the literature,  the SACs of various types of traffic processes have been found \cite{Jiang2008book, Jiang2010}. Examples are as follows. For a periodic arrival process with period length $T$, $\alpha(t)=\frac{L}{T}(t)+L$ and $f(x) = 0$. For a Poisson arrival process with rate $\lambda$, $\alpha(t) = \frac{\lambda t}{\theta}(e^{\theta L}-1)$ and $f(x) = e^{-\theta x}$ for any $\theta >0$. For a Markov On-Off arrival process, in which, the source generates no arrivals at state Off and produces constant rate traffic at state On with rate $r$, and the transition rate from state On to state Off is $\lambda$ and the transition rate from state Off to state On is $\mu$,  $\alpha(t) = \frac{1}{{2\theta }}\left( {\theta r - \lambda  - \mu  + \sqrt {{{(\theta r + \lambda  - \mu )}^2} + 4\lambda \mu } } \right)\cdot t$ and $ f(x)= e^{-\theta x} $. 

Assuming i.i.d. packet service times, the SSC has the following expressions \cite{Jiang2010}:  
\begin{eqnarray}
\beta(t) &=& R \cdot t \label{servicerate} \\
g(x) &=& \mathop e\nolimits^{- \theta \frac{x}{R}} \label{sscboundfun} 
\end{eqnarray}
with $R = \frac{L \cdot \theta }{\ln(M_\tau(\theta))}$ for $\theta >0$, where $L$ denotes the packet length and $M_\tau(\theta)$ the moment generating function (MGF) of packet service time $\tau$, i.e., $M_\tau(\theta)=E[ e^{\theta \cdot \tau }]$ for any $\theta>0$. 

With these SAC and SSC results, the corresponding stochastic delay bounds for different types of arrival process are readily obtained from Theorem \ref{SNCbounds}.

\nop{
\begin{defn}(Stochastic Arrival Curve)\label{SSA}
A flow is said to have a v.b.c (virtual backlog centric) stochastic arrival curve $\alpha(t)\in F$\footnotemark with bounding function
$f(x)\in \overline F$\footnotemark, if its arrival process $A(t)$ satisfies, for any $t\geq0$ and all $x\geq0$ there holds:
\begin{equation}\label{vbc SAC}
\Pr\{\sup_{0\leq s \leq t}[A(s,t)-\alpha(t-s)]> x\}\leq f(x).
\end{equation}
where $A(s,t)$ denotes the cumulative amount of traffic of the flow during period $(s, t]$, $A(t)= A(0,t)$, and $\alpha(t)$ is a non-decreasing function.
\end{defn}
\footnotetext{$F$: the set of non-negative wide-sensing increasing functions}
\footnote{$\overline F$: the set of non-negative wide-sensing decreasing functions}

In order to characterize the service process of a channel with error and retransmission, we present a stochastic service curve by using a virtual time function approach \cite{6662937} as:

\begin{thm}(Stochastic Service Curve)\label{SSC}
The considered system has a stochastic service curve $\beta(t) = R \cdot (t-\frac{L}{R})^+$,$\beta \in F$ with bounding function $g \in \overline F$, denoted by $S \sim ssc\left\langle {g,\beta } \right\rangle $, if for all $t\geq 0$ and all $x \geq 0$ there holds:
\begin{equation}\label{ssc}
\Pr\{A\otimes\beta(t)-A^\ast(t)> x\}\leq g(x),
\end{equation}
\begin{equation}\label{sscboundfun}
g(x) = \mathop e\nolimits^{- \theta \frac{x}{R}}
\end{equation}
\begin{equation}\label{servicerate}
R = \frac{L \cdot \theta }{\ln(M_\tau(\theta))}
\end{equation}
\end{thm}
where$A \otimes \beta (t) \equiv in{f_{0 \le s \le t}}\left\{ {A(s) + \beta (t - s)} \right\}$,${A^ * }(t)$ denote the cumulative output traffic amount up to time t and $M_\tau(\theta)$ denote the moment generating function(MGF) of packet service time $\tau$, i.e., $M_\tau(\theta)=E[ e^{\theta \cdot \tau }]$ for any $\theta>0$.
In order to obtain R we have to first determine the MGF of packet service time. Since the distribution of the packet service time is difficult to find out an explicit formulation so far. Therefore, we propose an empirical approximation approach \ref{sec:modelMGF} that allows us make use of the empirical model of Packet Error Rate(PER) to obtain the packet service time MGF later. Having defined SAC and SSC, the following performance bounds have been derived as follows\cite{Jiang2008book}:

\begin{thm}(Stochastic performance bounds)\label{SNCbounds}
	If a system provides a stochastic service curve $\beta(t)$ with bounding function $g(x)$ to a flow, which has v.b.c stochastic arrival curve $\alpha(t)$ with bounding function $f(x)$ , if $\beta(t)$ and $\alpha(t)$ are independent, then for any $t\geq0$ and all $x\geq 0$, the flow has a delay $D(t)$ bound as:
%	the flow has a delay $D(t)$ and backlog$B(t)$ bounds as:
%\begin{equation}\label{backlogbound}
%%P\{ B(t) > x\}  \le \bar F \otimes \bar G(x - \alpha \oslash \beta (0))
%P\left\{ {B(t) > x} \right\} \le 1 - \overline f  * \overline f (x - \alpha  \oslash \beta (0))
%\end{equation}

\begin{equation}\label{delaybound}
%P\{ D(t) > h(\alpha  + x,\beta )\}  \le \bar F \otimes \bar G(x)
P\left\{ {D(t) > h(\alpha  + x,\beta )} \right\} \le 1 - \overline f  * \overline g (x)
\end{equation}
%where $\alpha \oslash \beta (0) = su{p_{u \ge 0}}\left\{ {\alpha (u) - \beta (u)} \right\}$, $h(\alpha  + x,\beta ) = su{p_{s \ge 0}}\left\{ {\inf \left\{ {\tau  \ge 0:\alpha (s) + x \le \beta (s + \tau )} \right\}} \right\}$, 
where$\overline f (x) = 1 - {\left[ {f(x)} \right]_1}$\footnotemark, $\overline g(x) = 1 - {\left[ {g(x)} \right]_1}$ and $*$ is the Stieltjes convolution operation $\overline f  * \overline g (x) = \int\limits_0^x {\overline g } (x - y)d\overline f (y)$.
\end{thm}
\footnotetext{$[\cdot]_1$denotes $\max(\min(\cdot,1),0)$}
}

%\subsection{ The Empirical approximation of the per-packet service time MGF}\label{sec:modelMGF}
%\subsection{Finding stochastic service curve of the wireless link}\label{sec:modelMGF}
\subsection{Service modeling of the wireless link}\label{sec:modelMGF}
To apply SNC results to delay distribution analysis of the wireless link, the discussion above indicates that we need to find a stochastic service curve for the link. To this aim, equations (\ref{servicerate}) and (\ref{sscboundfun}) are the bridge, where a variable that needs to be decided is the rate $R$. 

In order to obtain $R$, we have to first find the MGF of packet service time. Since the distribution of the packet service time is unknown and difficult to obtain directly, we consider the following approach by making use of the empirical model for Packet Error Rate (PER) obtained from measurement. 

%Since the distribution of the packet service time is difficult to find out an explicit formulation so far. Therefore, we propose an empirical approximation approach \ref{sec:modelMGF} that allows us make use of the empirical model of Packet Error Rate(PER) to obtain the packet service time MGF later. 

As introduced in Sec.\ref{sec:emprical-mod}, the per-packet service/transmission time is determined by the random variable $n_{Tries}$, i.e., the number of transmissions per packet. The probability that a packet is not transmitted successfully is determined by PER $P_e$, which is also the packet retransmission probability. In contrast, for each transmission attempt the successful probability is $1 - P_e$. %As ~\figref{sysmodel} illustrated, 
As a result, $n_{Tries}$ has truncated geometric distribution with limit on the maximum number of transmissions $N_{maxTries}$ per packet, from which the distribution of per-packet service time $T$ is readily found. 

Specifically, we have %with unit transmission time per packet, the per-packet service time distribution $P(T)$ is given by:
\begin{eqnarray} 
%&&\Pr(T)=\Pr(n_{Tries})	\nonumber\\
&&\Pr(T)	\nonumber\\
&=&\left\{{\begin{array}{l}
	{\Pr({T_{ACK}})},(1\le{n_{tires}}\le{N_{maxtires}})\\
	{\Pr({T_{nonACK}})}, (otherwise)%(n_{tires}=N_{maxtires})
	\end{array}} \right.\nonumber\\
&=&\left\{{\begin{array}{l}
	(1-{P_e}){P_e}^{{n_{tires}}-1},(1 \le {n_{tires}} \le {N_{maxtires}})\\
	1-\sum\limits_{{n_{tires}}=1}^{{N_{maxtires}}}{(1-{P_e}){P_e}^{{n_{tires}}-1}}, (otherwise)%(n_{tires}=N_{maxtires})
	\end{array}} \right.\nonumber
\end{eqnarray}
where $P_e$ is approximated by (\ref{equ:PER}), and $T_{ACK}$ and $T_{nonACK}$ are given in (\ref{equ:servicetimeACK}) and (\ref{equ:servicetimenonACK}) respectively. Further, 
according to the definition of moment generation function, an approximation of packet service time MGF $ M_{T}(\theta)$ is given by:
%\[\left( {\sum\limits_{{n_{tries}} = 1}^{N_{maxtires}} } {{e^{\theta {T_{ACK}}}}P({T_{ACK}})} } \right) + {e^{\theta {T_{nonACK}}}}P({T_{nonACK}})\]
\begin{align}\label{TserviceMGF}
&M_{T}(\theta )= E(e^{T\theta }) \nonumber\\
&=\left(\sum\limits_{n_{tries}=1}^{N_{maxtires}} e^{\theta T_{ACK}}\Pr(T_{ACK})\right)+e^{\theta T_{\mathit{nonACK}}}\Pr(\mathit{T_{nonACK}})
\end{align}
where $\theta>0$ is in a small neighborhood of zero.

Applying (\ref{TserviceMGF}) to (\ref{servicerate}) and (\ref{sscboundfun}), a stochastic service curve of the link is obtained, based on which delay distribution analysis can be further conducted with Theorem \ref{SNCbounds}.

\begin{figure}[htbp]
	\centering
	\includegraphics[width=1.0\columnwidth]{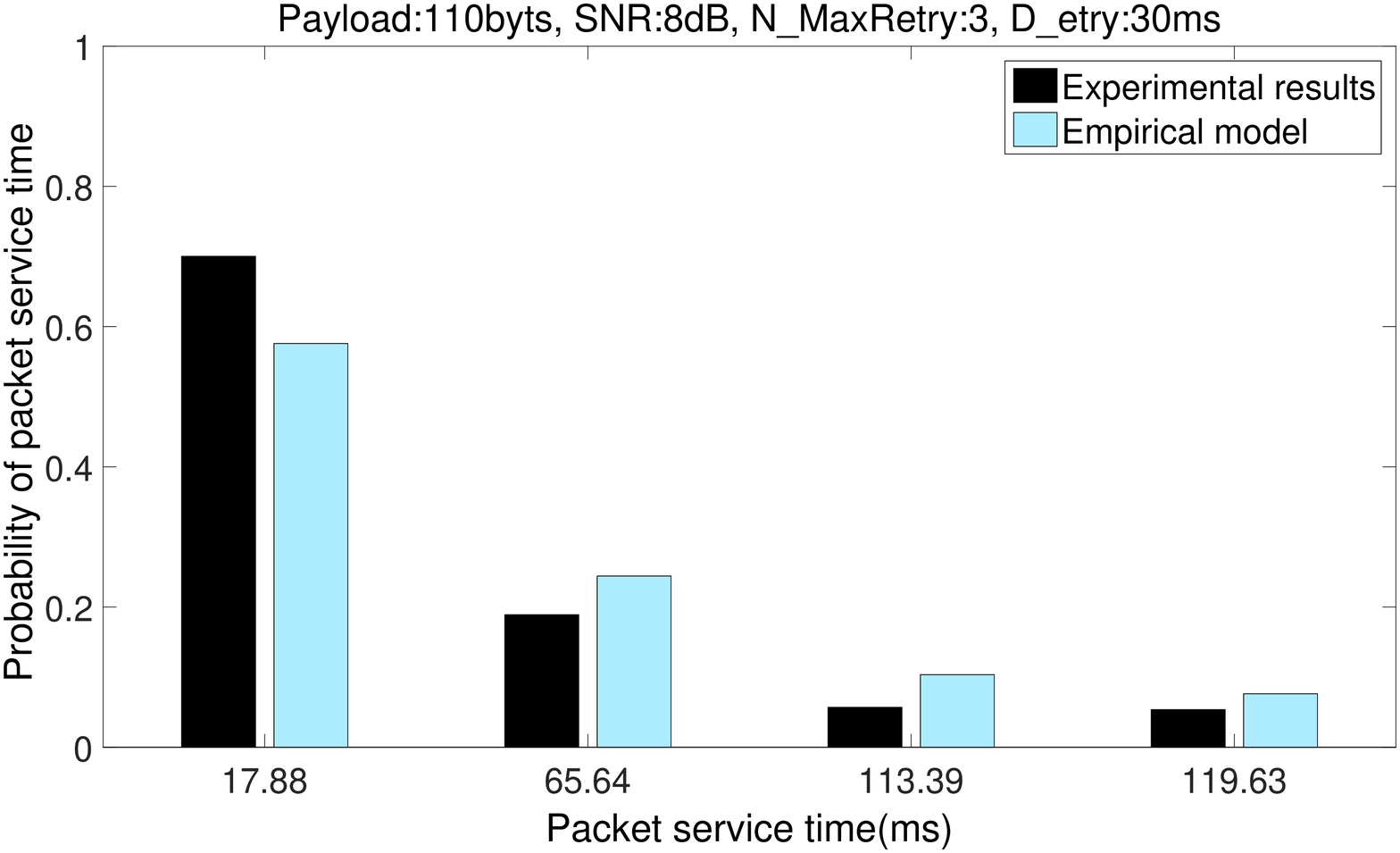}
	\caption{Validation of packet service time distribution model}
	\label{PrTsermodelvalid}
\end{figure}

\begin{figure}[htbp]
	\centering
	\includegraphics[width=1.0\columnwidth]{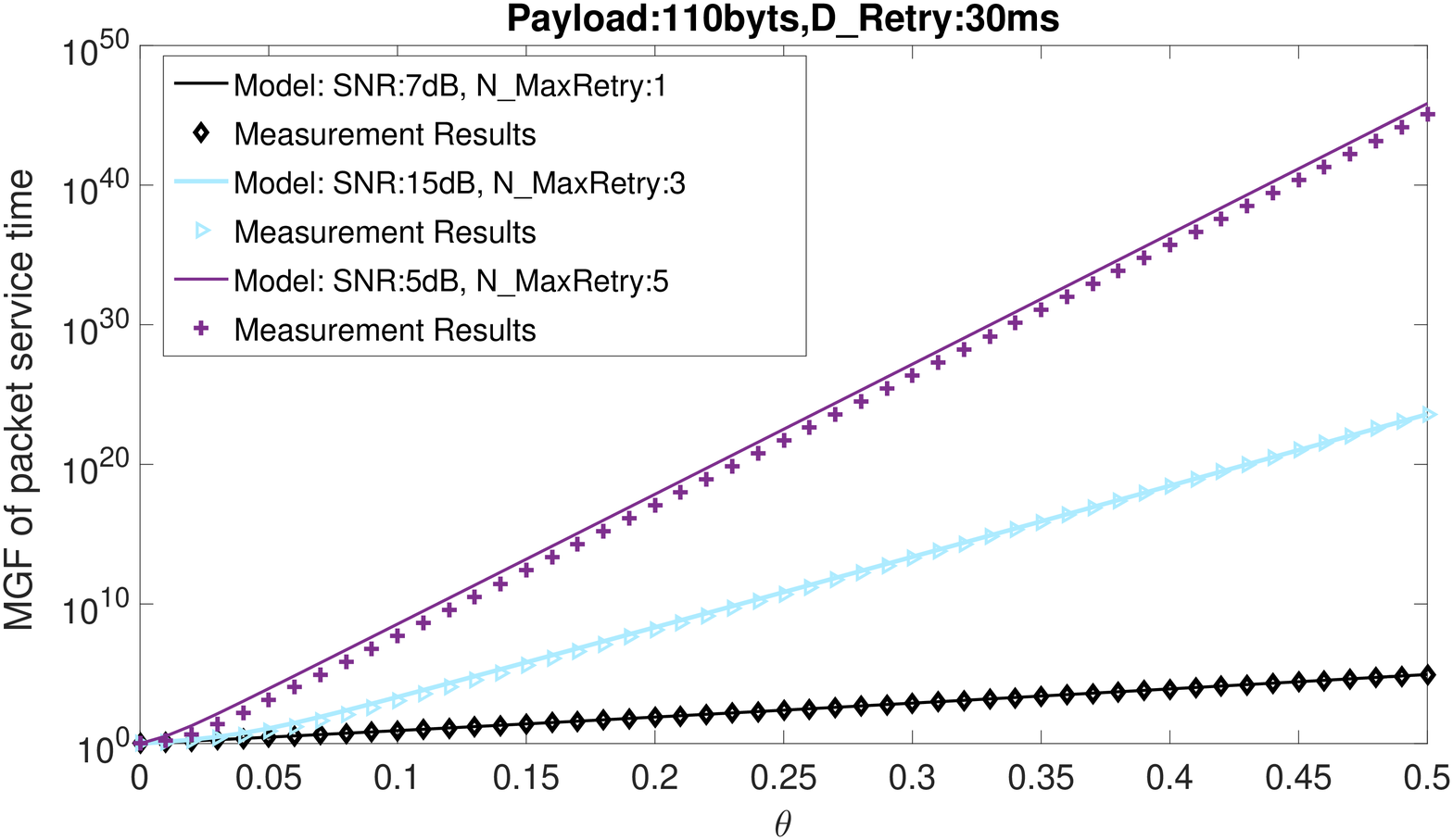}
	\caption{Validation of empirical model of packet service time MGF }
	\label{MGFmodelvalid}
\end{figure}

{\em Validation:} 
To verify the distribution and MGF empirical approximation model of packet service time, we analyzed the statistics of measured packet service time under different parameter settings. We compared results from the model (\ref{TserviceMGF}) with those from statistical analysis of the per-packet service time measured in different parameter configurations. As illustrated in ~\figref{PrTsermodelvalid}, the truncated geometric distribution matches the measurement data well, meaning that the distributions of packet service time can be approximately modeled by PER and truncated geometric distribution. Further, by using statistic analysis of the measured per-packet service time, the MGF of packet service time can be computed.  ~\figref{MGFmodelvalid} shows the comparison between computed MGF of measured packet service time and the proposed model as the different parameter setting.  As \figref{MGFmodelvalid} illustrated, the approximated MGF is in good agreement with the measured result. 
% As \figref{MGFmodelvalid} illustrated, for larger $SNR$ values, e.g. $SNR \geq 5dB$, the approximated MGF is close to the measurement results. Because when $SNR\leq 5dB$ the measured PER has large fluctuation and thus the PER fitting is not very well. However, in general the approximate MGF of packet service time model are in good agreement with the measured results.

%\subsection{Analytical results and empirical validation}\label{bounds}
\subsection{Model validation using delay distribution}\label{bounds}
In Section \ref{sec-mvm}, we have used mean delay to validate the equivalent $G/G/1/\infty$ queueing model for delay analysis. In the present subsection, we investigate the validity of the model using delay distribution. 

%In this section, we will use the empirical approximated stochastic service curve obtained by our indoor experiments for periodic traffic and Theorem \ref{SNCbounds} in sec.\ref{sec:SNCthm} to present the delay performance bounds under different traffic sources. Three traffic cases are considered: the periodic traffic pattern can be modeled by a stochastic arrival curve; Poisson traffic and Markov modulated traffic patterns can be modeled by a stochastic arrival curve. For all performance bounds in following figures, the free parameters $\theta $ are optimized numerically. To validate the analytical approximation model, the empirical evaluations are presented for delay distributions. The empirical experiments have been conducted on indoor wireless link under various parameters settings(as shown by Table~\ref{tab:ExpParameter}) .

%\subsubsection{Periodic traffic arrival}\label{sec:periodictraffic}
\nop{
 When packet flow arrival process is periodic with the same packet size $L$ and the inter-arrival time $T$, the packet arrival process $A(t)$ is bounded by a deterministic arrival curve$\alpha(t)$ where $\alpha(t)=rt+b=\frac{L}{T}(t)+L$.\cite{BoudecT01} According to the above stochastic service curve\ref{SSC} description, the delay performance bound can be readily derived as follows:
 \begin{proposition}\label{periodbounds}
 	When $\frac{L}{\tau} \le R$, for any time $t \ge 0$,
\begin{eqnarray}
&&\Pr\{(D(t) \geq \frac{L+x}{R}\} \leq  e^{- \theta \frac{x}{R}} \label{eq.periodicdelaybound} \\		 
%&&\Pr\{B_{pkt}(t)\geq L + x\} \leq e^{- \theta \frac{x}{R}}\label{eq.periodicbacklogbound}  \\
&&	\text{subject to }
	\theta>0 \text{ and } \frac{L}{T} \le R
\end{eqnarray} 
where  $R$ is given by (\ref{servicerate} ).
\end{proposition}
}

\begin{figure}[th]
	\centering
	\includegraphics[width=1.0\columnwidth]{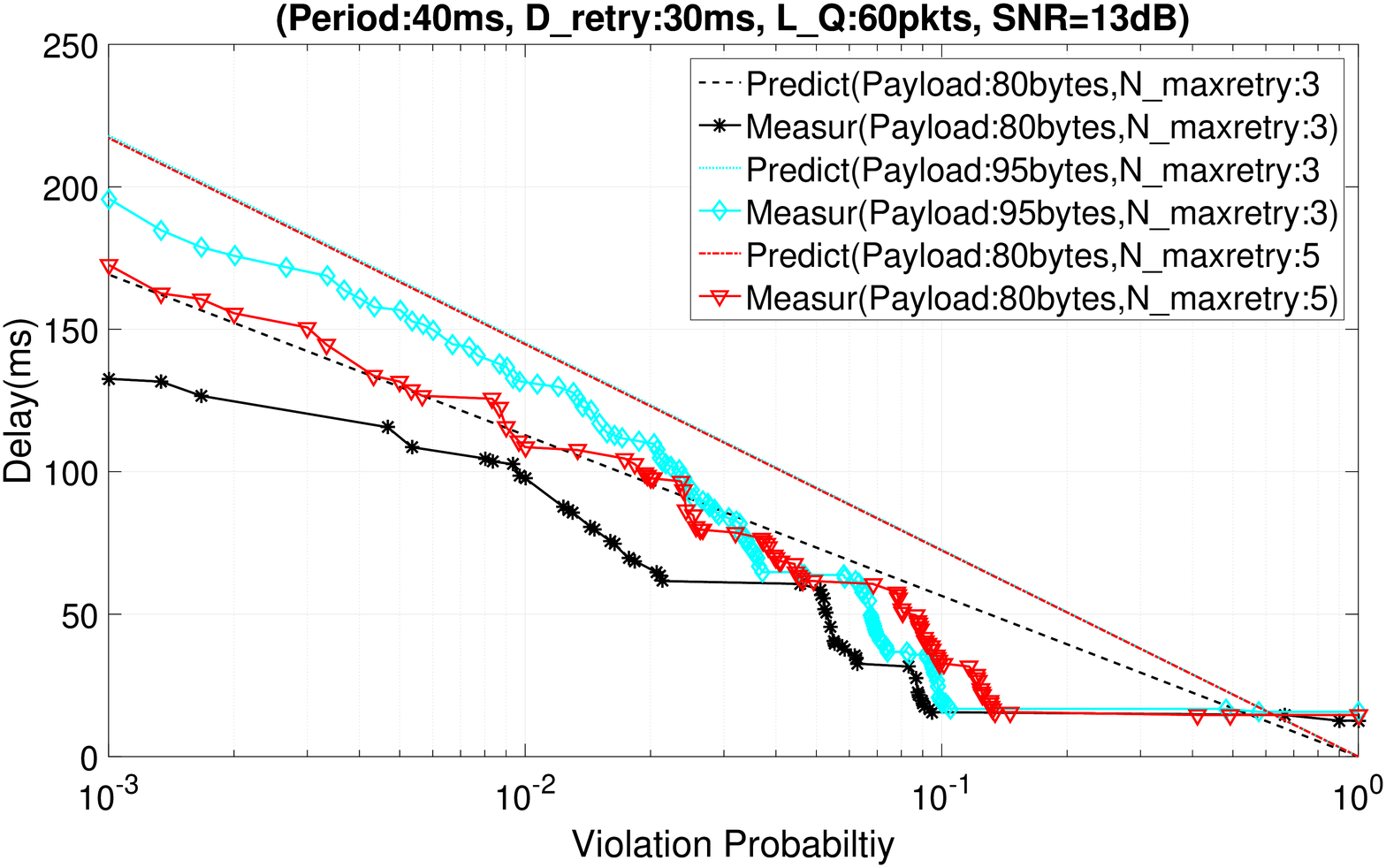}
	\caption{Validation under different parameter configurations}%{Validation under different parameter configurations: $L=80,95$, $SNR=13dB$, $T=40ms$,  $N_{\text{maxretry}}=3,5$, $D_{\text{retry}}=30ms$, $L_q=60$}
	\label{delaybound_periodic_fixsnr}
\end{figure}

\begin{figure}[th]
	\centering
	\includegraphics[width=1.0\columnwidth]{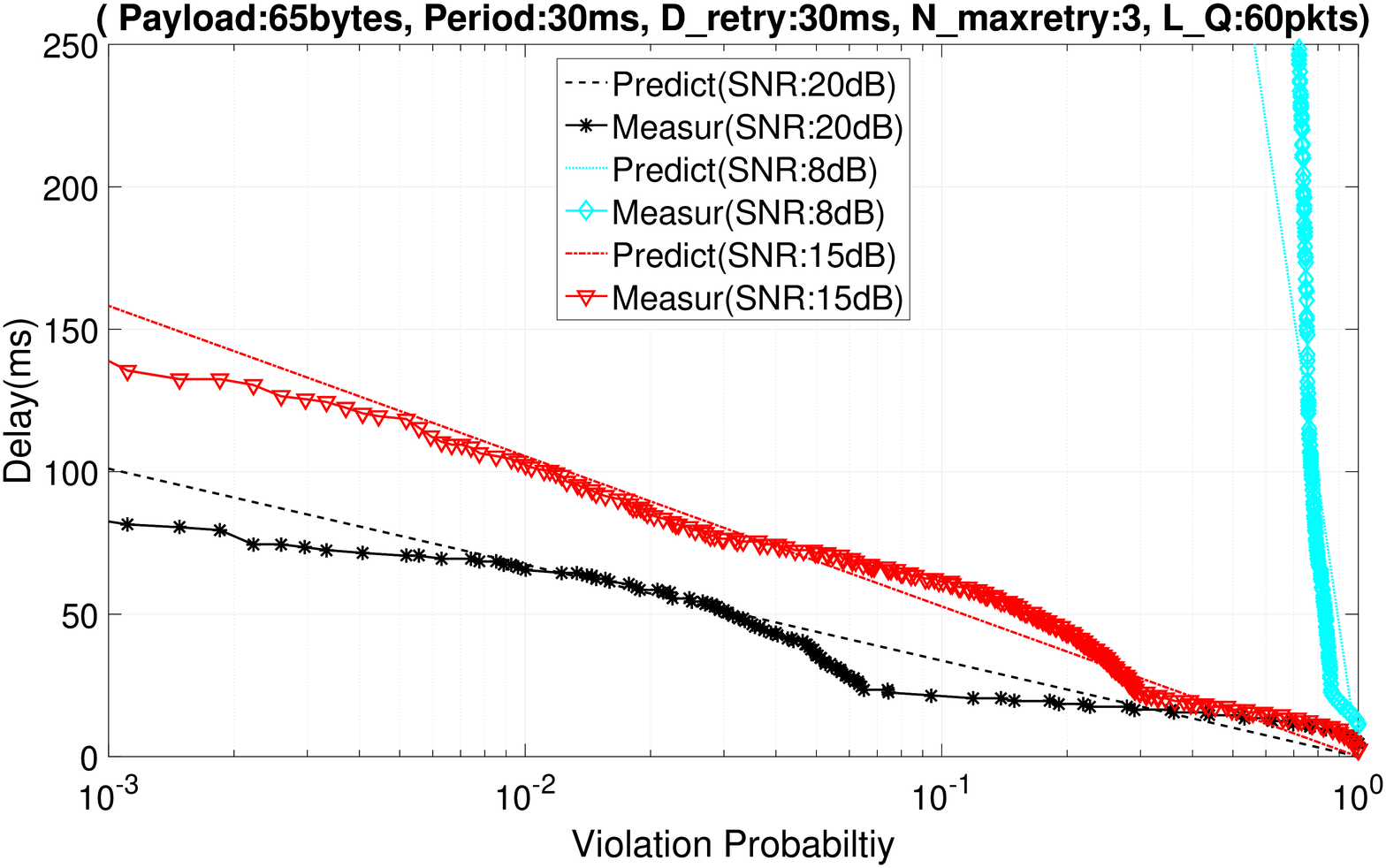}
	\caption{Validation under different SNRs}
%{Validation under different SNRs: $SNR=8,15,20dB$,  $L=65$, $T=30ms$, $N_{\text{maxretry}}=3$, $D_{\text{retry}}=30ms$, $L_q=60$}
	\label{delaybound_periodic_diffsnr}
\end{figure}

The results of validation are illustrated in ~\figref{delaybound_periodic_fixsnr} and ~\figref{delaybound_periodic_diffsnr}. %For the former, the $SNR=13dB$ and the payload length is $80,95$ bytes, the packet inter-arrival time $Period=40ms$, the maximum number of transmission tries is set to $N_{\text{maxretry}}=3,5$ with the retry delay $D_{\text{retry}}=30ms$ and the maximum queue size is set to $L_q=60$ packets. For the latter,   $SNR=8,15,20dB$ and the payload length is $65$ bytes, the packet inter-arrival time $Period=30ms$, the maximum number of transmission tries is set to $N_{\text{maxretry}}=3$ with the retry delay $D_{\text{retry}}=30ms$ and the maximum queue size is set to $L_q=60$ packets. 
As can be observed, the predicted distributions based on analysis agree well with the measurement results for all cases. A cause of the slight difference in the matching is the MGF approximation model of packet service time, which has been used in deriving SSC and delay distribution bound. The results also show that higher traffic load (larger payload size and shorter packet inter-arrival time) and smaller SNR lead to increase in the delay, which is also predicted by our model. In addition, ~\figref{delaybound_periodic_fixsnr} shows that by reducing the maximum number of transmission attempts $N_{\text{maxretry}}$ the delay becomes smaller but will lead to lowering the reliability. Furthermore, increasing $N_{\text{maxretry}}$ and reducing payload size together make the analytical delay distribution bounds closer to each other, which is also validated by experimental results. %This is due to the multi parameter effects. Therefore, to balance the service reliability and latency we can easily find the parameter configuration solution through our analytical approach. 

\subsection{Further results}\label{bounds2}

In the remaining, we show analytical delay distribution results based on the proposed equivalent queueing model, the empirical MGF based stochastic service curve characterization of the link and the stochastic network calculus theory, and compare with experimental results to conversely validate the analytical approximation model. Two additional traffic pattens are considered: Poisson traffic and Markov On-Off traffic. Their stochastic arrival curves have been given in Section \ref{sec:SNCthm}, with which, the corresponding delay distribution bounds are readily obtained from Theorem \ref{SNCbounds}. For all performance bounds in the following figures, the free parameters $\theta $ are optimized numerically. 

To validate the analytical approximation model, measurement results are also presented for delay distribution, for which additional experiments were conducted on the indoor wireless link using these traffic patterns. The distance between two nodes is 15m and the transmit power level for sender node is varied from 11 to 31. The packet delay for 3, 000 packets is recorded for each experiment. 

%we will use the empirical approximated stochastic service curve obtained by our indoor experiments for periodic traffic and Theorem \ref{SNCbounds} in sec.\ref{sec:SNCthm} to present the delay performance bounds under different traffic sources. Three traffic cases are considered: the periodic traffic pattern can be modeled by a stochastic arrival curve; Poisson traffic and Markov modulated traffic patterns can be modeled by a stochastic arrival curve. For all performance bounds in following figures, the free parameters $\theta $ are optimized numerically. To validate the analytical approximation model, the empirical evaluations are presented for delay distributions. The empirical experiments have been conducted on indoor wireless link under various parameters settings(as shown by Table~\ref{tab:ExpParameter}) .

\subsubsection{Poisson traffic}
 \begin{figure}[thb]
 	\centering
 	\includegraphics[width=1.0\columnwidth]{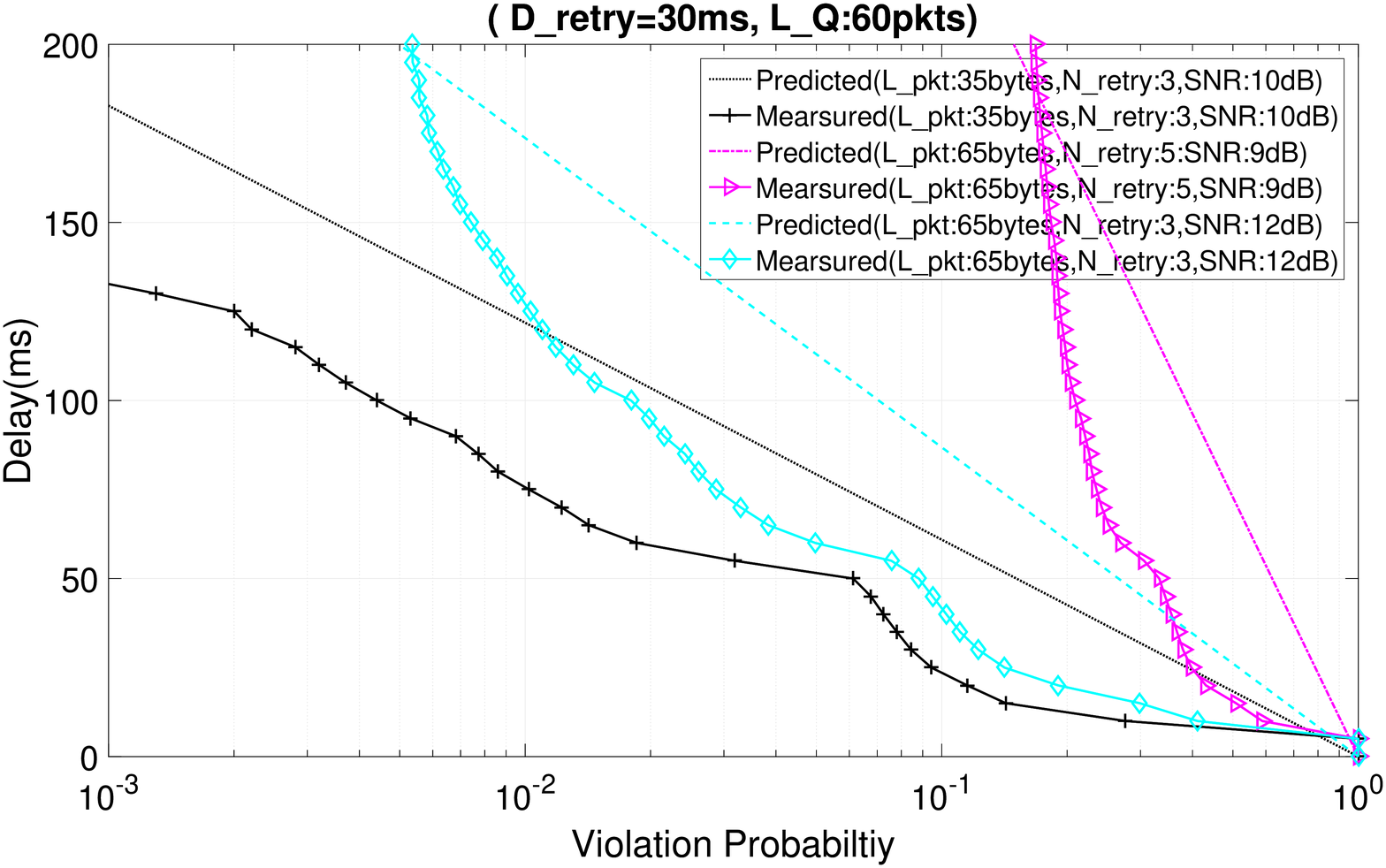}
 	\caption{Delay distribution under Poisson arrival: $\lambda=30\text{Packets/s}$.}
 	\label{fig:delaybound_Poisson}
 \end{figure}
 
   \begin{figure}[thb]
 	\centering
 	\includegraphics[width=1.0\columnwidth]{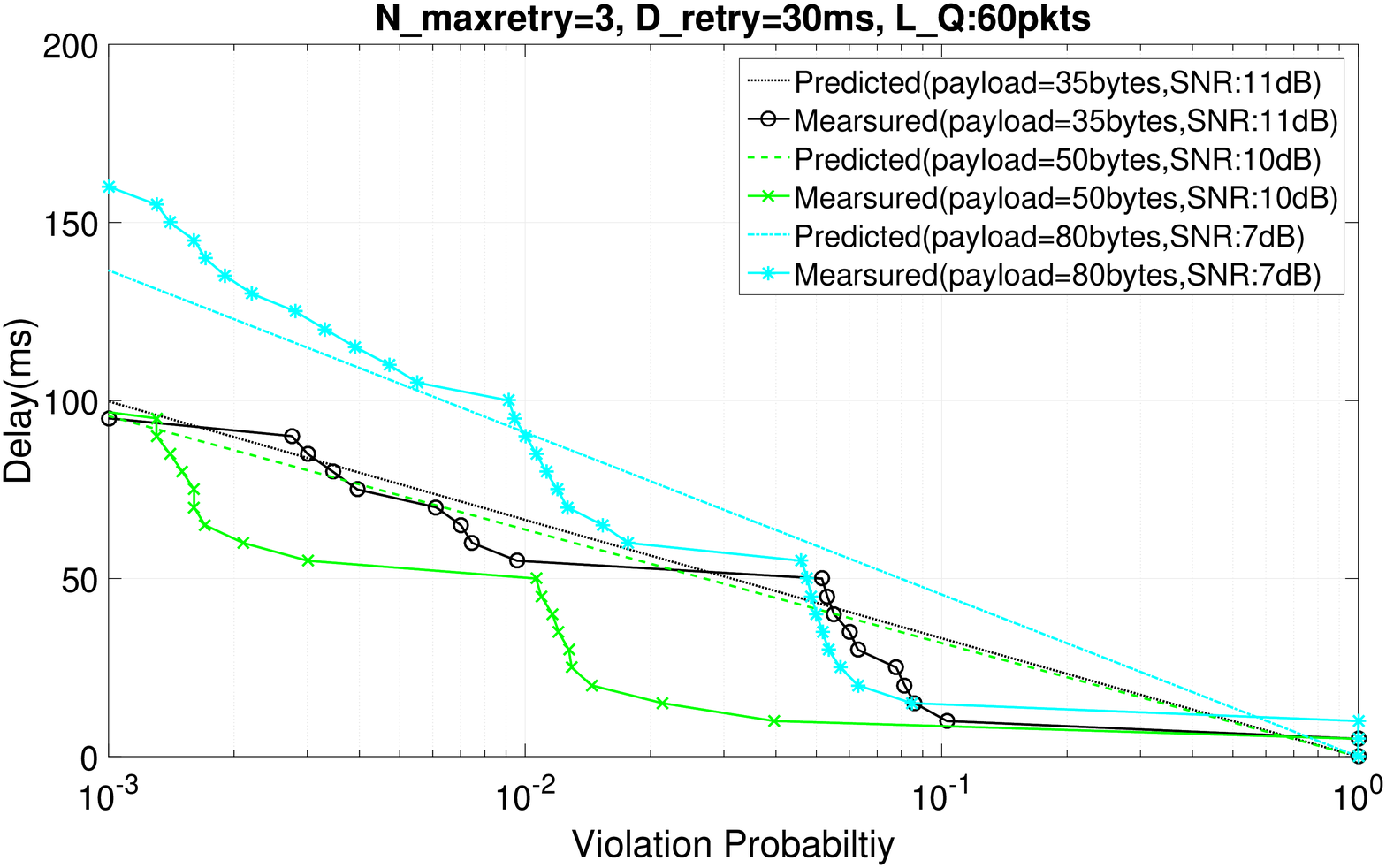}
 	\caption{Delay distribution under On-Off arrival}
 	\label{fig:delaybound_onoff}
 \end{figure}
 
~\figref{fig:delaybound_Poisson} shows the \textit{CCDF}s of measured delay along with the predictions by our analytical delay bounds. It shows that the predictions are closely along and in the same order of magnitude as the experiment results. It can be observed that the predictions, as analytical bounds, approximate with the measurements. The results shown in ~\figref{fig:delaybound_Poisson} also reveal that heavier traffic (larger payload) leads to longer delay and lower reliability as can be observed from the analysis. However, for the same traffic, by reducing the maximum number of transmission attempts$N_{\text{maxretry}}$ and increasing SNR, the delay decreases and reliability increases. Overall, ~\figref{fig:delaybound_Poisson}  shows that the delay performance under Poisson traffic can be effectively predicted, bounding the delay distribution, by our SNC-based analysis.

 \nop{
Poisson traffic is one of the most widely used and oldest traffic model in literature. The inter-arrival times are memoryless, i.e., exponential. We denote $A(t)$ the Poisson packet arrival process with average arrival rate $\lambda$ and all packets have the same size L. By using the effective bandwidth, the arrival traffic flow can be modeled as a v.b.c stochastic arrival curve $\alpha(t)$ with bounding function $f_{Poisson}(x)$ \cite{kelly1996stochastic}\cite{Chang2000} as:
 \begin{eqnarray}
 \alpha(t) &=& \frac{\lambda t}{\theta}(e^{\theta L}-1) \\
 f_{Poisson}(x) &=& e^{-\theta x} \label{eq:poissonbound} 
 \end{eqnarray}
 where $\theta>0$ satisfying $\alpha(t) \leq Rt$ and $R$ is given by \ref{servicerate}. 
 
By making use of the results from stochastic network calculus as reviewed in Sec. \ref{sec:SNCthm}, the following delay performance bound for the Poisson packet arrival can be expressed as follows:
 
 \begin{proposition}\label{poissionbounds}
 	\begin{eqnarray}
 	&&\Pr\{(D_{pkt}(t) \geq \frac{L +x}{R}\} \leq 1 - \overline f_{Poisson}  * \overline g (x-L) \\		 
% 	&&\Pr\{B_{pkt}(t)\geq L+ x\} \leq  1 - \overline f_{Poisson}  * \overline g (x)\\
 	&&\text{subject to }
 	\theta>0 \text{ and } \frac{L}{\theta}(e^{\theta L}-1) \le R
 	\end{eqnarray} 
 	where R is given by (\ref{servicerate} ) , $\overline f_{Poisson}(x)=1-f_{Poisson}(x)$ given by (\ref{eq:poissonbound}) and $\overline g(x)=1-g(x)$ given by (\ref{sscboundfun}).
 \end{proposition}
 }

 \subsubsection{Two state Markov On-Off traffic}
 
To further check the applicability of our SNC-based delay distribution analysis, another traffic pattern, Markov On-Off, is considered. In this Markov modulated packet arrival process, $\lambda=30$pkts/s, $\mu=20 $pkts/s and periodic packet arrival rate $r=20$pkts/s during state On. The results of both analytical and experiment results are illustrated in ~\figref{fig:delaybound_onoff}. The figure shows that the delay bounds of the analytical model approximatively agree with the experiment results. The results suggest that increasing transmit power (to achive higher SNR) and reducing the packet payload significantly improve the delay performance as also captured by the analytical evaluations. Overall, it can be observed that our analytical model also performs well under Markov on-off traffic. 
  
  \nop{
 At end, we consider a general traffic: Markov modulated arrival processes with a two state Markov chain. Markovian traffic is frequently used to model the On-Off characteristics of certain sources such as voice. In this case, the delay performance cannot be easily analyzed by applying traditional queuing theory. However the stochastic network calculus offers convenient methods to obtain the performance bounds.
 
 According to the On-Off model, the packet arrival process $A(t)$ has two states where the current state depends (only) on the previous state. In state Off (state 1) the source generates no arrivals, and in state On (state 2) it generates constant rate process with rate parameter $r$. The transition rate from state 2 to state 1 is $\lambda$ and the transition rate from state1 to state 2 is $\mu$. The Markov On-Off process has a v.b.c stochastic arrival curve 
 $ A(t){\sim_{sac}}\langle {f_{\mathit{Markov}}}, \alpha(t) \rangle $ for  $\forall \theta>0$ and 
 
 \begin{align}\label{eq:onoffcurve}
 \alpha(t)& =\rho (\theta )\cdot t\nonumber\\
 & = \frac{1}{{2\theta }}\left( {\theta r - \lambda  - \mu  + \sqrt {{{(\theta r + \lambda  - \mu )}^2} + 4\lambda \mu } } \right)\cdot t
 \end{align}
 
 with bounding function \cite{Jiang2008book}:
 
 \begin{equation}\label{eq:onoffbound} 
 f_{\mathit{Markov}}(x)= e^{-\theta x} 
 \end{equation}
where $\theta>0$ satisfying $\alpha(t) \leq R\cdot t$ and $R$ is given by \ref{servicerate}. 
 
 By making use of the results from Sec. \ref{sec:SNCthm}, the following delay performance bound for two state Markov On-Off arrival traffic can be expressed as follows:
 
 \begin{proposition}\label{eq:onoffperfbounds}
 	\begin{eqnarray}
 	&&\Pr\{(D_{pkt}(t) \geq \frac{L+x}{R}\} \leq 1 - \overline f_{\mathit{Markov}}  * \overline g (x-L) \\		 
% 	&&\Pr\{B_{pkt}(t)\geq L+ x\} \leq  1 - \overline f_{\mathit{Markov}} * \overline g (x)\\
 	&&\text{subject to }
 	\theta>0 \text{ and } \frac{L}{\theta}(e^{\theta L}-1) \le R
 	\end{eqnarray} 
 	where R is given by (\ref{servicerate} ) , $\overline f_{\mathit{Markov}}=1-f_{\mathit{Markov}}$ given by (\ref{eq:onoffbound}) and $\overline g(x)=1-g(x)$ given by (\ref{sscboundfun}).
 \end{proposition}
 
 \begin{figure}[thb]
 	\centering
 	\includegraphics[width=1.0\columnwidth]{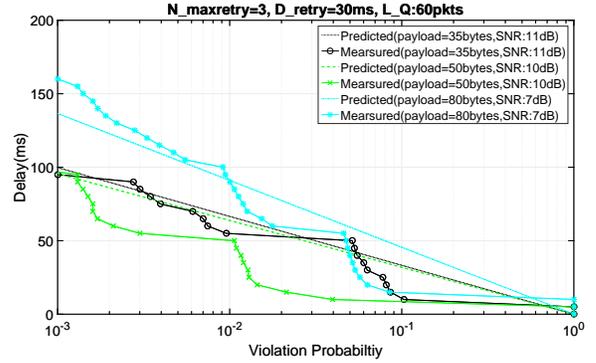}
 	\caption{Delay distribution under On-Off arrival}
 	\label{fig:delaybound_onoff}
 \end{figure}
 }

\section{Conclusion} \label{sec:conclusion} % and future work
In this paper, we investigated the performance of packet delivery over a point-to-point 802.15.4 wireless link. Exploiting measurement results, two methods, respectively based on the classical queueing theory and the newly established stochastic network calculus (SNC) theory, have been introduced to estimate / predict the delay performance. Specifically, to ease delay analysis, an equivalent $G/G/1$ non-loss queueing model is constructed. Then, a classical queueing theory result is utilized to estimate the mean delay performance. Afterwards, SNC is applied to estimate the delay distribution performance, for which, a method to model the service time is introduced. In all these, i.e., queueing model, mean delay analysis, service time modeling and delay distribution analysis, measurement results have played crucial roles. They have not only been analyzed to establish empirical models that are utilized in these steps, but also been compared with results from the analysis, under a wide range of settings. The comparisons, showing a good match between analytical results and measurement results, give a clear indication on the effectiveness and validity of the proposed equivalent $G/G/1$ model, the classical queueing theory based mean delay analysis and the SNC based delay distribution analysis. 

%In this paper, we characterized the data delivery schemes over a point-to-point 802.15.4 wireless link. Based on the measurement results, we provide two queuing analysis methods to predict the delay performance. Firstly, we have established an equivalent G/G/1 non-loss queuing model for our given traffic pattern in experiments by using queuing theory. Secondly, making use of our available experiment data we provide a practical probabilistic analysis approach to predict the packet delay performance. Combining empirical model, we conduct stochastic service curve to model the packet service time then obtain the delay performance bound which can be applicable to many types of traffic pattern. This proposed approach has been validated by testbed experiments for different traffic pattern. Moreover, in the potential real-world wireless application the analytic approach allow us to identify and give the guidelines for setting the multi-layer parameters to achieve the QoS performance guarantee of wireless link. As a future research, we will our analytical model to capture more different transmission scenarios. We also plan to using our performance prediction models to on-line optimal and adjust WSNs link system parameter configurations.  

%\bibliographystyle{bibtex/IEEEtran}
%\bibliography{bibtex/sabrinabib}

\flushend

\end{document}